\documentclass[prb,aps,amsmath,showpacs,twocolumn]{revtex4}
\usepackage{graphicx}

\newcommand{\ds}{\displaystyle}

\renewcommand{\Im}{\mathop{\rm Im}}
\renewcommand{\Re}{\mathop{\rm Re}}

\begin{document}

\title{Coherent-potential-approximation study of excitonic absorption in orientationally disordered
molecular aggregates}

\author{D.B. Balagurov}
\email{d.balagurov@sns.it}
\author{G.C. La Rocca}
\affiliation{Scuola Normale Superiore and INFM, Piazza dei
Cavalieri 7, 56126 Pisa, Italy}
\author{V.M. Agranovich}
\affiliation{Institute of Spectroscopy, Russian Academy of
Sciences, 142190 Troitsk, Moscow Region, Russia}

\begin{abstract}

We study the dynamics of a single Frenkel exciton in a disordered
molecular chain. The coherent-potential approximation (CPA) is
applied to the situation when the single-molecule excitation
energies as well as the transition dipole moments, both their
absolute values and orientations, are random. Such model is
believed to be relevant for the description of the linear optical
properties of one-dimensional $J$ aggregates. We calculate the
exciton density of states, the linear absorption spectra and the
exciton coherence length which reveals itself in the linear
optics. A detailed analysis of the low-disorder limit of the
theory is presented. In particular, we derive asymptotic formulas
relating the absorption linewidth and the exciton coherence length
to the strength of disorder. Such expressions account
simultaneously for all the above types of disorder and reduce to
well-established form when no disorder in the transition dipoles
is present. The theory is applied to the case of purely
orientational disorder and is shown to agree well with exact
numerical diagonalization.

\end{abstract}

\pacs{71.35.Aa; 78.30.Ly}

\maketitle

\section{\label{sec:Introd} Introduction}

The optically active states in organic molecular
aggregates~\cite{Jelley,KnoesterRev}
--- the spatially-regular linear arrangements of dye
molecules --- are one-dimensional Frenkel excitons.~\cite{Davydov}
The nature of the excitonic states to be extended over many
molecules reveals itself in the narrowing of the absorption
resonances and the shortening of the radiative lifetime upon
aggregate formation (effects known as exchange narrowing and
``superradiance").~\cite{Knapp} The related coherence length
coincides with the total number of molecules in the aggregate only
if the size of the latter is sufficiently small. Otherwise various
exciton decoherence mechanisms, among which the static disorder is
the most essential at low temperature, substantially reduce the
coherence length that becomes independent from the actual size of
the aggregate. Considerable work has been done to analyze the
aspects of disorder in the molecular excitation
energies~\cite{Knapp,MalyshevScal,FidderWiersma,Knoester,Adame,MalMoreno,Makhov,MalMal}
and positions.~\cite{Fidder,MalyshevPoz,Kauffmann,Shimizu,Avgin}
Much less attention has been devoted to the disorder in
orientations~\cite{Markovitsi} or, generally, in the transition
dipole moments of individual molecules. However, from the limited
available information on the actual structure of molecular
arrangements one cannot exclude that the transition dipoles are
seriously influenced by disorder. Moreover, the last type of
disorder affects in a nontrivial way the separate components of
the optical susceptibility tensor and, hence, would be observable
in polarization-resolved absorption and luminescence
experiments\cite{PolarRes}. The present work has been motivated by
the need for a detailed theoretical analysis of orientational
disorder, as well as clarification of certain aspects concerning
the exciton coherence length in molecular aggregates.

In this paper the linear optical properties of a disordered
molecular chain are studied using the single-site
coherent-potential approximation~\cite{Soven,RevEl} (CPA). Applied
to various disorder problems, such as that of the electronic
structure in a random alloy, this self-consistent approximation
was shown to reproduce well the single-particle characteristics
like the density of states (DOS). As for the systems in which
elementary excitations are Frenkel excitons, the CPA is in
addition capable to provide the complete information on the linear
optical response, because the latter is extracted only from the
single-exciton Green's function (GF) averaged over disorder
realizations. The CPA has been successfully used to model the
optical spectra of periodic molecular arrangements with random
on-site energies~\cite{Huber1,Huber2} and
composition.~\cite{BakalisKnoester} However, if the transition
dipoles are also affected by disorder, a proper modification of
the theory is required. The key observation which allows us to do
this is that the dipoles enter the off-diagonal part of the
Hamiltonian in a bilinear form. In this case a single-site
approximation can be constructed according to a vector analog of
the Shiba ansatz.~\cite{Shiba} The resulting scheme is equivalent
to the matrix extension of CPA derived by Blackman, Esterling and
Berk~\cite{Blackman} (BEB) for compositionally-disordered alloys
with random hopping energies. [The analytical properties of BEB
CPA have been examined in
Refs.~\onlinecite{EsterSolo,GonisGarland,Kopernik,BEBStruct}].
Persson and Liebsch~\cite{Persson}, and Rozenbaum {\it et
al.}~\cite{Rozenbaum} have constructed a similar version of CPA to
study, respectively, the susceptibility of polarizable particles
and the vibrational modes of coupled oscillators randomly and
isotropically oriented in two or three dimensions. However, their
theory is still not applicable to the molecular aggregates in
which one typically deals with an essentially anisotropic
distribution of the transition dipoles.

The paper is organized as follows. In the next section we define
the model of a disordered molecular aggregate and set up the basic
formalism for the forthcoming calculations. The exciton Green's
tensor is introduced to express the related single-particle
quantities: the DOS, the linear polarizability, and the coherence
length. In Sec.~\ref{sec:CPA} we present the general formulation
of the BEB-CPA scheme and in Sec.~\ref{sec:WeakDis} derive its
analytic solution in the low-disorder limit. In
Sec.~\ref{sec:Results} the theory is applied to a specific case of
orientational disorder and compared to the results of the exact
numerical diagonalization.

\section{\label{sec:Formalism} Basic formalism}

\subsection{\label{sec:Model} Model}

An aggregate can be considered as a chain of ${\cal N}$ identical
two-level molecules. The linear optical properties in the
resonance region are determined by the single-particle states of
the Frenkel Hamiltonian~\cite{Davydov}
\begin{equation}
\label{Frenkel} H = \sum_n \epsilon_n B^\dagger_n B_n + \sum_{n,m}
J_{nm} B^\dagger_n B_m.
\end{equation}
The operator $B^\dagger_n$ ($B_n$) creates (annihilates) an
electronic excitation of energy $\epsilon_n$ on the $n$th molecule
(whose excited state is supposed to be nondegenerate). The
transfer terms $J_{nm}$ result from the Coulomb interaction
between an excited molecule and one in the ground state.
Accounting only for the dipole-dipole contribution one has
\begin{equation}
\label{TransferTerm} J_{nm} = \sum_{\alpha,\beta} p_n^\alpha
\vartheta^{\alpha\beta}_{nm} p_m^\beta,
\end{equation}
where $p_n^\alpha$ are the vector components of the $n$th molecule
transition dipole. The ``coupling kernel''
$\vartheta^{\alpha\beta}_{nm}$ carries information on the
dielectric function of the surrounding material and the location
of the molecules, but not on their dipole moments. If the dynamics
of the exciton subsystem is nondissipative (and the molecules have
no magnetic structure, so that $p^\alpha_n$ can be chosen real)
$\vartheta^{\alpha\beta}_{nm}$ has to be symmetric in the
tensorial indices ($\vartheta^{\alpha\beta}_{nm} =
\vartheta^{\beta\alpha}_{nm}$). We shall assume the molecules to
form a one-dimensional regular lattice such that
$\vartheta^{\alpha\beta}_{nm}$ depends solely on the
intermolecular distance $|n - m|$ (measured in the units of
lattice spacing). The theory will be presented in a general way
without specifying a concrete form of this dependence, i.e., for
any screening and anisotropy of the Coulomb interaction that can
occur in a solvent or on a substrate interface.

The disorder enters in our model both through the on-site energies
$\epsilon_n$ and the dipole moments $p^\alpha_n$. We shall assume
the families of random parameters $\{ \epsilon_n, p_n^\alpha \}$
corresponding to distinct sites to be mutually independent and to
have identical probability distribution. At the same time we still
allow for correlations between the transition energy and the
dipole components of a given molecule.

Finally, we shall neglect everywhere the effects of a finite
length of the aggregate considering the limit ${\cal N} \to
\infty$. This approximation is justified for sufficiently strong
disorder when the exciton coherence length found for the infinite
system does not exceed the actual length of the chain.

\subsection{\label{sec:Basic} Exciton Green's tensor}

For a given random realization the total information on the
aggregate's optical dynamics is imbedded in the complete set of
single-exciton eigenstates $\sum_n \psi_n(E) B_n^\dagger \left| 0
\right>$; the vacuum $\left| 0 \right>$ is the direct product of
the molecular ground states; $\psi_n(E)$ is the real-space exciton
wave function with an eigenenergy $E$. The local density of states
(LDOS) on the $n$th site and the total (normalized) exciton DOS
are found as
\begin{equation}
\label{RhoNdef} \rho_n(\omega) = \sum_{ \{ E \} } |\psi_n(E)|^2
\delta(\omega - E) = -\frac1{\pi} \Im G_{nn}(\omega),
\end{equation}
\begin{equation}
\label{RhoDef} \rho(\omega) = \frac1{\cal N} \sum_{ \{ E \} }
\delta(\omega - E) = -\frac1{\pi {\cal N}} \sum_n \Im
G_{nn}(\omega),
\end{equation}
where the real-space single-exciton GF is given by
\begin{equation}
\label{ScalarGreen}
\begin{array}{l}
\ds G_{nm}(z) = -i \int_0^{+\infty} d t \, e^{i z t} \langle 0 |
B_n(t) B_m^\dagger(0) | 0 \rangle
\\
\\
\ds \phantom{G_{nm}(z)} = \sum_{ \{ E \} } \frac{ \psi_n(E)
\psi^*_m(E)}{z - E}
\end{array}
\end{equation}
and the symbol $\{ E \}$ indicates that summation is performed
over all eigenstates. Here and below the GF for a real-valued
argument $\omega$ is found as the limit $z \to \omega + i0^+$.
Assuming that around the single-exciton resonance the light
wavelength is much larger than the spatial extent of a typical
excitonic wave function the aggregate interacts with the optical
field as a pointlike dipole. Hence, the linear response to the
external optical field is completely described by the
polarizability tensor
\begin{equation}
\chi^{\alpha\beta}(z) = i \int_0^{+\infty} dt \, e^{i z t} \langle
0 | [P^\alpha(t), P^\beta(0)] | 0 \rangle.
\end{equation}
in which the operator of total polarization represents the sum of
local polarizations, $P^\alpha = \sum_n \bigl( P^{\alpha (+)}_n +
P^{\alpha(-)}_n \bigr)$, both positive- and negative-energy
components:
\begin{equation}
\label{PolarizOper} P^{\alpha (+)}_n = p^\alpha_n B^\dagger_n,
\qquad P^{\alpha (-)}_n = p^\alpha_n B_n.
\end{equation}
To set up a convenient formalism for the forthcoming discussion
let us define the time-ordered two-point correlator of the local
polarizations
\begin{equation}
\label{GMatr} \Gamma^{\alpha\beta}_{nm}(z) = -i \int_0^{+\infty} d
t \, e^{i z t} \langle 0 | P^{\alpha (-)}_n(t) P^{\beta (+)}_m(0)
| 0 \rangle,
\end{equation}
or, equivalently,
\begin{equation}
\label{TensToScal}
\Gamma^{\alpha\beta}_{nm}(z) = p^\alpha_n G_{nm}(z) p^\beta_m.
\end{equation}
The newly introduced quantity will be referred through the paper
as the Green's tensor (GT), as opposed to the ``scalar'' GF
considered above. Combining the previous formulas we arrive at the
straightforward relation
\begin{equation}
\label{AbsDef} \chi^{\alpha\beta}(z) = -\sum_{n,m} \left[
\Gamma^{\alpha\beta}_{nm}(z) + \Gamma^{\alpha\beta *}_{nm}(-z^*)
\right],
\end{equation}
which indicates that (up to the sign factor) the GT coincides with
the positive-energy counterpart of the local linear response
function.

As usual one is interested not in the solution of the entire
dynamical problem for a given disorder realization, but in finding
the configurational average of some basic observable parameters.
In our case the quantities of interest are those defined in
Eqs.~(\ref{RhoDef}) and (\ref{AbsDef}). The subsequent
implementation of CPA suggests that the disorder-averaged DOS
$\bar \rho(\omega) \equiv \left< \rho(\omega) \right>$ and
polarizability $\bar \chi^{\alpha\beta}(z) \equiv \left<
\chi^{\alpha\beta}(z) \right>$ are to be expressed in terms of the
conditionally-averaged on-site GT
\begin{equation}
\label{CAGamma} \tilde \Gamma^{\alpha\beta}_{nn}(z) \equiv \left<
\Gamma^{\alpha\beta}_{nn}(z) \right>_{{\rm all \,\, sites \,\,
except} \,\, n{\rm th}},
\end{equation}
with fixed $n$th site variables $\{\epsilon_n, p^\alpha_n \}$, and
the complete statistical average
\begin{equation}
\label{AGamma} \bar \Gamma^{\alpha\beta}_{nm}(z) \equiv \left<
\Gamma^{\alpha\beta}_{nm}(z) \right>,
\end{equation}
the only quantities accessible within a single-site effective
theory. To proceed with the DOS it is sufficient to notice that,
because the $n$th-site dipoles $p_n^\alpha$ remain fixed, the
conditional average of the local GT and GF are still related to
each other as established in Eq.~(\ref{TensToScal}), $\tilde
\Gamma^{\alpha\beta}_{nn}(z) = p^\alpha_n \tilde G_{nn}(z)
p^\beta_n$. As a result the scalar GF is obtained via projection
\begin{equation}
\tilde G_{nn}(z) = \sum_{\alpha,\beta} \frac{p^\alpha_n}{|p_n|^2}
\tilde \Gamma^{\alpha\beta}_{nn}(z) \frac{p^\beta_n}{|p_n|^2},
\quad |p_n|^2 = \sum_\alpha (p^\alpha_n)^2.
\end{equation}
Using this relationship the conditionally-averaged LDOS can be
expressed as
\begin{equation}
\label{DALRho} \tilde \rho_n(\omega) = -\frac1{\pi} \Im
\sum_{\alpha,\beta} \frac{p^\alpha_n}{|p_n|^2} \tilde
\Gamma^{\alpha\beta}_{nn}(\omega) \frac{p^\beta_n}{|p_n|^2},
\end{equation}
while the total DOS is found in terms of $\tilde
\Gamma_{nn}^{\alpha\beta}(z)$ as a trivial single-site average
\begin{equation}
\label{DARho} \bar \rho(\omega) = -\frac1{\pi} \Im
\sum_{\alpha,\beta} \left< \frac{p^\alpha_n}{|p_n|^2} \tilde
\Gamma^{\alpha\beta}_{nn}(\omega) \frac{p^\beta_n}{|p_n|^2}
\right>.
\end{equation}
Remarkably, the statistical averaging in the last formula can be
performed immediately if the probability distribution of the
transition dipoles is of the ``purely orientational'' form. The
trace over vector indices of the conditionally-averaged GT depends
only on the deterministic, in this case, absolute value of the
$n$th-site dipole moment $|p_n|$. The averaged GT and GF are
related to each other as
\begin{equation}
\sum_\alpha \bar \Gamma^{\alpha\alpha}_{nn}(z) = |p_n|^2 \bar
G_{nn}(z),
\end{equation}
and consequently,
\begin{equation} \label{OrientDOS} \bar
\rho(\omega) = -\frac1{\pi |p_n|^2} \sum_\alpha \Im \bar
\Gamma_{nn}^{\alpha\alpha}(\omega).
\end{equation}

The straightforward application of the averaging procedure to
Eq.~(\ref{AbsDef}) gives the disorder-averaged linear
polarizability in the form
\begin{equation}
\label{DAChi} \bar \chi^{\alpha\beta}(z) = -{\cal N}\sum_m \left[
\bar \Gamma^{\alpha\beta}_{nm}(z) + \bar \Gamma^{\alpha\beta
*}_{nm}(-z^*) \right].
\end{equation}
In order to reduce the double real-space summation we have used
the translational invariance of $\bar
\Gamma^{\alpha\beta}_{nm}(z)$ that allows to replace in both terms
of Eq.~(\ref{DAChi}) $\sum_{n,m} \bar
\Gamma^{\alpha\beta}_{nm}(z)$ by ${\cal N} \sum_m \bar
\Gamma^{\alpha\beta}_{nm}(z)$. The last sum here is already
converging and does not depend on its upper limit ${\cal N}$.
Equivalently,
\begin{equation}
\label{DAChiFour} \bar \chi^{\alpha\beta}(z) = - {\cal N} \left[
\bar \Gamma^{\alpha\beta}_{k=0}(z) + \bar \Gamma^{\alpha\beta
*}_{k=0}(-z^*) \right],
\end{equation}
where the momentum-domain disorder-averaged GT is introduced
according to
\begin{equation}
\label{GFouMatr} \bar \Gamma^{\alpha\beta}_{nm}(z) =
\int_{-\pi}^{\pi} \frac{d k}{2\pi} \, e^{i k (n - m)} \bar
\Gamma^{\alpha\beta}_k(z),
\end{equation}
$k$ being measured in units of the inverse lattice spacing.
Noticeably, the above polarizability scales linearly with the
number ${\cal N}$ of molecules constituting the aggregate. Unlike
the physics coming from the nontrivial dependence of $\bar
\Gamma^{\alpha\beta}_{nm}(z)$ on the energy variable $z$ and the
intersite separation $|n-m|$ this elementary aspect is not related
to the degree of exciton coherence. Nevertheless it guarantees
fulfillment of an important part of the general sum rule according
to which the polarizability is an extensive quantity proportional
to the total number of polarizable objects.

\subsection{\label{sec:CohLength} Coherence length}

In the literature dealing with the optics of molecular aggregates
it is usually introduced the notion of the exciton coherence
length~\cite{MalMal,Fidder,Shimizu,Markovitsi} to characterize the
spatial extension of the exciton wave functions. The cited works,
addressing the problem mainly with numerical methods, employ
several definitions of this parameter to be extracted from the set
of wave functions obtained with explicit diagonalization
procedures. Even though the quantitative estimates provided within
all approaches can be in reasonable agreement with each other, the
physical arguments used are somewhat different. From the viewpoint
of the present work a natural definition of the exciton coherence
length can be given on the basis of Eq.~(\ref{DAChi}). In fact,
entering it the disorder-averaged single-exciton GT contains all
the information on the linear optical response of the excitonic
system and depends nontrivially on the amplitude and phase
coherence between the wave functions for different disorder
realizations. These coherence properties partly reveal themselves
in the DOS and absorption spectra which can be essentially
different from those of the disorder-free system, but it is also
the dependence of $\bar\Gamma^{\alpha\beta}_{nm}(z)$ on the
intersite separation that gets strongly affected by the structural
disorder within the chain. Excluding some peculiar
cases,~\cite{Kozlov} each component of the disorder-averaged GT is
characterized by an exponential behavior
\begin{equation}
\label{NCohDef} \bar \Gamma^{\alpha\beta}_{nm}(z) \sim \exp \!
\left( i \xi^{\alpha\beta}(z) |n - m| \right), \quad |n - m| \to
\infty,
\end{equation}
where $\xi^{\alpha\beta}(z)$ is a complex-valued wave number.
Keeping in mind this asymptotics while performing the real-space
summation in Eq.~(\ref{DAChi}) one concludes that the dominant
contribution to the disorder-averaged polarizability
$\bar\chi^{\alpha\beta}(z)$ comes from the pairs of sites with
$\Im \xi^{\alpha\beta}(z)|n - m| \lesssim 1$. The parameter
\begin{equation}
\label{NCD1} N^{\alpha\beta}(z) = \frac1{\Im
\xi^{\alpha\beta}(z)},
\end{equation}
indicating how many molecules contribute to a given component of
the linear polarizability, provides a natural definition of the
exciton coherence length. Clearly, $N^{\alpha\beta}(z)$ is a
straightforward generalization of the conventional quasiparticle
phase coherence length, defined in terms of the
large-intersite-separation asymptotics of the GF, to incorporate
the vectorial nature of the excitonic polarization. The quantity
(\ref{NCD1}) can be also thought of as the nonlocality range of
the linear optical response function.

\section{\label{sec:CPA} BEB-CPA scheme}

\subsection{\label{sec:CPAGeneral} Main procedure}

As already mentioned, in order to find the statistical averages
(\ref{CAGamma}) and (\ref{AGamma}) we shall employ the single-site
self-consistent approximation known as BEB CPA.\cite{Blackman}
This theory is capable to address simultaneously both diagonal and
off-diagonal disorder, provided the second enters the random
Hamiltonian in a generalized multiplicative form. The last
condition is intrinsically fulfilled in the excitonic problem
under consideration because the coupling (\ref{TransferTerm}) is
given by a bilinear combination of the random $p^\alpha_n$ with
$\vartheta^{\alpha\beta}_{nm}$ being translationally-invariant
deterministic matrix. The BEB extension of the scalar
CPA~\cite{Soven} is based on the so-called BEB
transformation.\cite{Blackman} The latter represents a vector
generalization of the multiplicative ansatz originally implemented
by Shiba~\cite{Shiba} to construct a single-site self-consistent
theory for bond-disordered alloys with mean-geometric relationship
between hopping integrals. Upon such transformation the problem of
finding the resolvent of a Hamiltonian with both diagonal and
multiplicative off-diagonal disorder is equivalently reformulated
for an operator which acts in a space with additional tensorial
dimensions but contains only site-diagonal disorder. With
application to Frenkel excitons the BEB transformation has been
already realized in Sec.~\ref{sec:Basic} by considering instead of
operators $B^\dagger_n$ ($B_n$) the polarizations
$P^{\alpha(+)}_n$ ($P^{\alpha(-)}_n$) equipped with vector
indices. As a consequence the dynamical equation for the scalar
GF,
\begin{equation}
\label{GreenMain} G_{nm}(z) = g_n(z) \delta_{nm} + g_n(z) \sum_l
J_{nl} G_{lm}(z),
\end{equation}
is replaced with that for the GT,
\begin{equation}
\label{GNJMatr} \Gamma_{nm}(z) = \gamma_n(z) \delta_{nm} +
\gamma_n(z) \sum_l \vartheta_{nl} \Gamma_{lm}(z).
\end{equation}
[Here and below omitting the indices we assume the usual rules for
multiplication and inversion of tensorial quantities.] Unlike the
case of Eq.~(\ref{GreenMain}), the disorder enters
Eq.~(\ref{GNJMatr}) only in the site-diagonal form, namely,
through the tensor
\begin{equation}
\label{G0Matr}
\gamma^{\alpha\beta}_n(z) = p^\alpha_n g_n(z) p^\beta_n
\end{equation}
associated with the bare local GF
\begin{equation}
\label{PropG} g_n(z) = \frac1{z - \epsilon_n}.
\end{equation}
At the same time, because the algebraic structure of equation
(\ref{GNJMatr}) remains similar to that of (\ref{GreenMain}), the
usual CPA self-consistency arguments can be employed to
approximate the statistical averages defined in
Eqs.~(\ref{CAGamma}) and (\ref{AGamma}). The BEB-CPA
scheme\cite{Blackman} is accomplished via straightforward
generalization of the scalar-CPA equations within the tensorial
formalism as outlined below.

The disorder effect on the single-particle quantities is accounted
in CPA by replacing the random local GTs
$\gamma^{\alpha\beta}_n(z)$ with a deterministic site-independent
coherent-potential GT $\gamma^{\alpha\beta}(z)$. The
disorder-averaged GT in the momentum representation is found from
Eq.~(\ref{GNJMatr}) as
\begin{equation}
\label{CPA2Matrix}
\bar \Gamma_k(z) = \left[ 1 - \gamma(z)
\vartheta_k \right]^{-1} \gamma(z),
\end{equation}
where
\begin{equation}
\label{TauKMatr} \vartheta^{\alpha\beta}_k = \frac1{\cal N}
\sum_{n,m} e^{-ik(n - m)} \vartheta^{\alpha\beta}_{nm}
\end{equation}
is the momentum-space coupling kernel. In turn, the
conditionally-averaged on-site GT $\tilde
\Gamma^{\alpha\beta}_{nn}(z)$ is approximated by assuming that the
site, carrying its random parameters, is placed into the same
coherent environment as the one of Eq.~(\ref{CPA2Matrix}). Thus,
finding this conditional average reduces to solving a
single-impurity problem realized by Eq.~(\ref{GNJMatr}) in which
the bare GTs $\gamma_m^{\alpha\beta}(z)$ for $m \ne n$ are
replaced by $\gamma^{\alpha\beta}(z)$. A simple
derivation~\cite{RevEl,Blackman,EsterSolo,GonisGarland} leads to
\begin{equation}
\label{GSkobN} \tilde \Gamma_{nn}(z) = \left[ 1 - \gamma_n(z)
\Sigma(z) \right]^{-1} \gamma_n(z),
\end{equation}
where the site-diagonal self-energy $\Sigma^{\alpha\beta}(z)$
describes coupling of the selected molecule with the
coherent-background molecules constituting the rest of the chain.
Clearly, being a functional of only the coherent-potential GT
$\gamma^{\alpha\beta}(z)$, this self-energy is the same as the one
entering the complete disorder-average of the local GT
\begin{equation}
\label{BarGSE} \bar \Gamma_{nn}(z) = \left[ 1 - \gamma(z)
\Sigma(z) \right]^{-1} \gamma(z).
\end{equation}
Provided both $\gamma^{\alpha\beta}(z)$ and $\bar
\Gamma^{\alpha\beta}_{nn}(z)$ are represented by nonsingular
matrices one gets
\begin{equation}
\label{SigmaSimple} \Sigma(z) = \gamma^{-1}(z) - \bar
\Gamma^{-1}_{nn}(z),
\end{equation}
where the local GT is found in terms of $\gamma^{\alpha\beta}(z)$
from Eq.~(\ref{CPA2Matrix}) after the momentum-space integration:
\begin{equation}
\label{CPA2Matr} \bar \Gamma_{nn}(z) = \int_{-\pi}^{\pi} \frac{d
k}{2\pi} \left[ 1 - \gamma(z) \vartheta_k \right]^{-1} \gamma(z).
\end{equation}

The unknown coherent-potential GT $\gamma^{\alpha\beta}(z)$ is to
be determined in a self-consistent way. In spirit of the original
CPA, one demands the single-site expectation of the
conditionally-averaged GT to coincide with the completely-averaged
local GT:
\begin{equation}
\label{CPA1Matr} \bar \Gamma^{\alpha\beta}_{nn}(z) = \bigl< \tilde
\Gamma^{\alpha\beta}_{nn}(z) \bigr>.
\end{equation}
The latter guarantees that the conditional and the complete
average of the GT found above will provide identical estimates for
the single-site quantities such as LDOS or the local
(single-molecule) polarizability.

Summarizing, the BEB-CPA procedure amounts to solving equations
which, including auxiliary definitions, are listed in formulas
from (\ref{G0Matr}) to (\ref{CPA1Matr}). For nontrivial disorder
models and realistic forms of the coupling
$\vartheta^{\alpha\beta}_{nm}$ this can be done only with the use
of numerical methods. An example of such numerical solution is
given in Sec.~\ref{sec:Results} [see also Appendix \ref{BEBExpl}].
The analytic treatment of the low-disorder limit of the theory  is
presented in Sec.~\ref{sec:WeakDis}.

\subsection{\label{sec:Anal} Analyticity and accuracy of BEB CPA}

As clear from the general discussion of Sec.~\ref{sec:Basic}, the
BEB CPA allows to find configurational averages of the exciton DOS
and linear polarizability of a disordered molecular aggregate. It
is natural to check the validity of such nonperturbative theory
asking, in particular, how close are the estimates to the actual
quantities and whether they meet certain fundamental physical
requirements. This question can be partly answered by analyzing
the structure of the BEB-CPA equations as done in a number of
publications.~\cite{RevEl,Blackman,EsterSolo,GonisGarland,Kopernik}
For completeness, let us outline the important facts concerning
analyticity and accuracy of the BEB CPA.

Gonis and Garland~\cite{GonisGarland} have proved that
$\bar\Gamma^{\alpha\beta}_{nm}(z)$ found within BEB CPA possesses
the same properties as a function of the complex energy $z$ as
would have the disorder-averaged GT calculated exactly. Namely, it
is analytic in the whole plane excluding branch cuts on the
semi-axis $\Im z = 0$, $\Re z > 0$, while the tensor
\begin{equation}
\label{SDensity} \bar A^{\alpha\beta}_k(z) = -\frac1{2\pi i}
\left[ \bar \Gamma^{\alpha\beta}_k(z) - \bar \Gamma^{\beta\alpha
*}_k(z) \right],
\end{equation}
that at $z = \omega + i0^+$ provides the exciton spectral density,
is positively (negatively) defined at $\Im z > 0$ ($\Im z < 0$).
As a consequence the disorder-averaged polarizability will
preserve causality, while the DOS and absorption will be
nonnegative.

The authors of Ref.~\onlinecite{Blackman} have studied the
accuracy of the theory to reproduce the energy-domain moments of
the spectral density, each given by a coefficient in the
high-energy expansion of the GT:
\begin{equation}
\label{LZAss} \bar\Gamma^{\alpha\beta}_k(z) = \sum_{s=0}^\infty
\frac{L^{\alpha\beta}_{k,s}}{ z^{s+1}}, \quad
L^{\alpha\beta}_{k,s} = \int_0^{+\infty} d \omega \, \omega^s \bar
A^{\alpha\beta}_k(\omega).
\end{equation}
It was shown that the BEB CPA provides correctly the first three
moments of the spectral density,
\begin{equation}
\label{Moms}
\begin{array}{c}
\ds L_{k,0} = l_0, \qquad L_{k,1} = l_1 + l_0 \vartheta_k l_0,
\phantom{\int\limits_{\pi}\!\!\!}
\\
\ds L_{k,2} = l_2 + l_1 \vartheta_k l_0 + l_0 \vartheta_k l_1 +
l_0 \vartheta_k l_0 \vartheta_k l_0,
\end{array}
\end{equation}
where the corresponding moments entering the high-energy expansion
of the local coherent-potential GT $\gamma^{\alpha\beta}(z)$ are
given by
\begin{equation}
\label{GamMom}
\begin{array}{c}
\ds l^{\alpha\beta}_0 = \left< p^\alpha_n p^\beta_n \right>,
\qquad l^{\alpha\beta}_1 = \left< p^\alpha_n \epsilon_n p^\beta_n
\right>, \phantom{\int\limits_{\pi}\!\!\!}
\\
\ds
\begin{array}{l}
\ds l^{\alpha\beta}_2 = \left< p^\alpha_n \epsilon^2_n p^\beta_n
\right> + \sum_{\gamma,\delta} \left[ \left< p^\alpha_n p^\gamma_n
p^\delta_n p^\beta_n \right> - \left< p^\alpha_n p^\gamma_n
\right> \left< p^\delta_n p^\beta_n \right> \right]
\\
\ds \phantom{ l^{\alpha\beta}_2 =} \times \int_{-\pi}^\pi
\frac{dk}{2\pi} \sum_{\mu,\nu} \vartheta^{\gamma\mu}_k \left<
p^\mu_n p^\nu_n \right> \vartheta^{\nu\delta}_k.
\end{array}
\end{array}
\end{equation}
The fact that the number of these moments is high enough
guarantees fulfillment of the sum rules for the DOS and
polarizability, correctly accounting for redistribution of the
total oscillator strength between components of the polarizability
tensor in the presence of orientational disorder. It follows also
that using BEB CPA one gets an exact value of the absorption
linewidth, {\it provided} the latter is defined from the weighted
second centered moment of the spectral density
[$\eta^{\alpha\beta} \sim \sqrt{L^{\alpha\beta}_{k=0,2}
L^{\alpha\beta}_{k=0,0} -
(L^{\alpha\beta}_{k=0,1})^2}/L^{\alpha\beta}_{k=0,0} $]. This
means that, even though there exists no unique way to define the
width of an inhomogeneously-broadened absorption line, the BEB CPA
still gives a reasonable estimate for this characteristics of the
spectrum.

Let us also mention that the theory does not seem to violate the
fundamental inequalities $|E - \epsilon_n| \le \sum_m |J_{nm}|$ to
be fulfilled for every eigenvalue $E$. These, in particular,
impose the lower, $\min(\epsilon_n - \sum_m |J_{nm}|)$, and the
upper, $\max(\epsilon_n + \sum_m |J_{nm}|)$, bounds of the
spectral region, where minimum and maximum are taken over all
disorder realizations. Even though no rigorous proof of this
property is known for the case of BEB CPA, it was shown to hold in
the numerical solution of the BEB-CPA equations for some random
alloy models.~\cite{Kopernik} As will be demonstrated in
Sec.~\ref{sec:Results}, these spectral bounds are not violated in
the case of orientational disorder either.

\begin{figure}
\centering
\includegraphics[height=4cm]{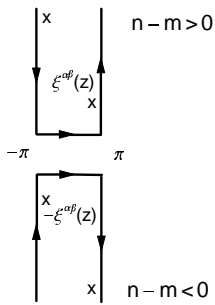} \caption{\label{Fig1} Schematic
picture of the integration contour in the complex momentum plane.
Each couple of vertical lines with opposite orientations give zero
contribution to the integral. The crosses mark poles of the
momentum-space GT.}
\end{figure}

Another important aspect of the theory is the behavior of the
disorder-averaged GT in the complex momentum domain. Considering
$\bar \Gamma^{\alpha\beta}_k(z)$ for complex $k$ provides
additional understanding about the exciton coherence length. Since
$\bar \Gamma^{\alpha\beta}_k(z)$ remains the same as $\Re k$ is
shifted by integer of $2\pi$, the integration contour in
Eq.~(\ref{GFouMatr}) can be modified as shown in Fig.~\ref{Fig1}.
Provided the function is meromorphic (i.e., has only isolated
poles) in $k$, the integration reduces to sum over poles in the
upper (lower) half-plane for $n-m > 0$ ($n-m < 0$). [Due to the
mirror symmetry of the problem each pole has a counterpart located
symmetrically with respect to the origin $k=0$.] To establish
connection with the coherence length defined above it is
sufficient to notice that only a single pair of poles will
contribute to the asymptotics (\ref{NCohDef}), namely, having
minimal distance from the real axis $\Im k = 0$. Applying these
arguments to the disorder-averaged GT in the BEB-CPA form
(\ref{CPA2Matrix}) one concludes that the complex wave numbers
$\xi^{\alpha\beta}(z)$ of Eq.~(\ref{NCohDef}) are to be selected
among the solutions $k$ of the characteristic equation
\begin{equation}
\label{CharEq} \det \left[ 1 - \gamma(z) \vartheta_k \right] = 0.
\end{equation}
Here we assumed that the Fourier-transformed kernel
$\vartheta^{\alpha\beta}_k$ is analytic in $k$ to provide
meromorphic $\bar \Gamma^{\alpha\beta}_k(z)$. This assumption is
valid, for instance, if the coupling extends only over a finite
number of sites. However, the last equation may not lead to a
correct coherence length for a more general nonanalytic coupling.

\section{\label{sec:WeakDis} Low-disorder limit}

For a disorder-free chain the GT obtained from the BEB CPA
coincides with the one known exactly. In order to establish the
relation of our approach with the existing scaling theories of
disordered one-dimensional excitons~\cite{MalyshevScal} let us
solve the BEB-CPA equations in the limit of weak disorder. A
similar calculation for the bond-disordered binary alloy has been
already done in Ref.~\onlinecite{Kopernik}. However, due to the
rather involved structure of BEB CPA the general treatment of the
low-disorder regime is still missing. For the sake of simplicity
we shall also relax generality imposing certain symmetries under
which the theory is formulated in terms of diagonal tensors.

\subsection{\label{sec:WDGen} General}

To proceed one needs first to specify the ``low" disorder in terms
of its probability distribution, entering the BEB CPA through the
self-consistency equation (\ref{CPA1Matr}). We do this imposing
smallness of the centered moments of the probability density which
characterize deviations of the on-site energies and transition
dipoles from their average values $\epsilon = \left< \epsilon_n
\right>$, $p^\alpha = \left< p^\alpha_n \right>$. The
conditionally-averaged local GT in Eq.~(\ref{CPA1Matr}) can be
represented as a power series in $\epsilon_n - \epsilon$ and
$p^\alpha_n - p^\alpha$, so that its $n$th-site expectation can be
written as
\begin{equation}
\label{LDMain} \bigl< \tilde \Gamma^{\alpha\beta}_{nn}(z) \bigr> =
\sum_\nu \bigl< \tilde \Gamma^{\alpha\beta}_{nn}(z) \bigr>_\nu.
\end{equation}
A $\nu$th term in the last sum is proportional to the
$\nu$th-order centered moments of the probability density. The
first three terms are given explicitly by
\begin{equation}
\label{MesExp}
\begin{array}{l}
\ds \bigl< \tilde \Gamma_{nn}(z) \bigr>_0 = \left. \tilde
\Gamma_{nn}(z) \right|_{\epsilon_n = \epsilon, p^\alpha_n =
p^\alpha},
\\
\\
\ds \bigl< \tilde \Gamma_{nn}(z) \bigr>_1 = 0,
\\
\\
\ds \!\! \begin{array}{l} \ds \bigl< \tilde \Gamma_{nn}(z)
\bigr>_2 = \frac12 \left[ a_2 \frac{\partial^2 \tilde
\Gamma_{nn}(z) }{\partial\epsilon_n^2} + \sum_\alpha b_2^\alpha
\frac{\partial^2 \tilde \Gamma_{nn}(z)}{\partial \epsilon_n
\partial p^\alpha_n} \right.
\\
\ds \phantom{\bigl< \tilde \Gamma_{nn}(z) \bigr>_2=} \left. \left.
+ \sum_{\alpha,\beta} c_2^{\alpha\beta} \frac{\partial^2 \tilde
\Gamma_{nn}(z)}{\partial p^\alpha_n
\partial p^\beta_n} \right] \right|_{\epsilon_n = \epsilon,
p^\alpha_n = p^\alpha},
\end{array}
\end{array}
\end{equation}
where in the last equation
\begin{equation}
\label{ABCDef}
\begin{array}{c}
\ds a_2 = \left< (\epsilon_n - \epsilon)^2 \right>, \qquad
b_2^\alpha = \left< (\epsilon_n - \epsilon)(p^\alpha_n - p^\alpha)
\right>,
\\
\ds c_2^{\alpha\beta} = \left< (p^\alpha_n - p^\alpha)(p^\beta_n -
p^\beta) \right> \phantom{\int\limits^\pi \!\!\!\!}
\end{array}
\end{equation}
are the second-order centered moments. The terms not included in
Eq.~(\ref{MesExp}) will be neglected following our assumption
about weakness of disorder. Later it will be demonstrated that the
information contained in the third- and higher-order centered
moments is indeed not preserved in the low-disorder limit of BEB
CPA.

We shall additionally impose some simplifying restrictions on the
form of disorder and intermolecular coupling to be considered
below. Namely, let the system to be such that (i) in some
(orthogonal) basis of the three-dimensional vector space the
coupling $\vartheta^{\alpha\beta}_{nm}$ is diagonal in the upper
indices for all pairs of molecules, and (ii) the disorder
probability density is symmetric under reflections of two basis
vectors. We shall label tensorial components corresponding to
either these two basis vectors by symbols ``$\perp_1$'',
``$\perp_2$'', and those of the remaining (third) vector by
``$\parallel$''. The coupling energies $J_{nm}$ and, hence, the
scalar GF $G_{nm}(z)$ remain invariant, while every off-diagonal
tensorial component of the GT $\Gamma^{\alpha\beta}_{nm}(z)$
changes its sign under one of the above reflections [because from
$\alpha \ne \beta$ it follows that either $\alpha$ or $\beta$ is
different from $\parallel$]. Therefore, configurational averaging
with a symmetric probability distribution will result to a
diagonal $\bar \Gamma^{\alpha\beta}_{nm}(z)$.

Regarding the BEB-CPA theory, from Eqs.~(\ref{CPA2Matrix}) and
(\ref{BarGSE}) it follows consecutively that both the
coherent-potential GT $\gamma^{\alpha\beta}(z)$ and the
self-energy $\Sigma^{\alpha\beta}(z)$ reduce to the diagonal form.
As concerns the low-disorder limit to be studied here, with our
symmetry requirements the average dipole moment $p^\alpha$ is
directed strictly along the $\parallel$-axis, the vectorial
parameter $b_2^\alpha$ defined in Eq.~(\ref{ABCDef}) has nonzero
component only along the same axis, while the tensor
$c_2^{\alpha\beta}$ is diagonal. Using the explicit dependence of
the conditionally-averaged GT on the $n$th site random parameters
[see Eqs.~(\ref{G0Matr}), (\ref{PropG}) and (\ref{GSkobN})] the
nontrivial (diagonal) components of the tensors (\ref{MesExp}) can
be found as
\begin{equation}
\label{dP0G} \left< \tilde \Gamma^\parallel_{nn}(z) \right>_0 =
\frac{|p|^2}{z-\epsilon - \Sigma^\parallel(z)|p|^2}, \quad \left<
\tilde \Gamma^{\perp_i}_{nn}(z) \right>_0 = 0,
\end{equation}
\begin{equation}
\label{dP1}
\begin{array}{l}
\begin{array}{l}
\ds \left< \tilde \Gamma^\parallel_{nn}(z) \right>_2 = \frac{a_2
|p|^2}{\bigl[z-\epsilon - \Sigma^\parallel(z)|p|^2\bigr]^3}
\\
\ds \phantom{\left< \tilde \Gamma^\parallel_{nn}(z) \right>_2 =} +
\frac{2 b_2^\parallel |p|}{\bigl[z-\epsilon -
\Sigma^\parallel(z)|p|^2\bigr]^2}
\\
\ds \phantom{\left< \tilde \Gamma^\parallel_{nn}(z) \right>_2 =} +
\frac{c_2^\parallel (z-\epsilon)\bigl[z-\epsilon + 3
\Sigma^\parallel(z)|p|^2\bigr]}{\bigl[z-\epsilon -
\Sigma^\parallel(z)|p|^2\bigr]^3}
\\
\ds \phantom{\left< \tilde \Gamma^\parallel_{nn}(z) \right>_2 =} +
\sum_{i=1,2} \frac{c_2^{\perp_i}
\Sigma^{\perp_i}(z)|p|^2}{\bigl[z-\epsilon -
\Sigma^\parallel(z)|p|^2\bigr]^2},
\end{array}
\\
\\
\ds \left< \tilde \Gamma^{\perp_i}_{nn}(z) \right>_2 =
\frac{c_2^{\perp_i}}{z - \epsilon - \Sigma^\parallel(z)|p|^2}.
\end{array}
\end{equation}
Combined with the BEB-CPA self-consistency equation
[Eq.~(\ref{CPA1Matr})] these formulas provide the starting point
for the analytical treatment of the low-disorder regime.

As seen from Eq.~(\ref{dP1}), the solution behaves quite
differently in the two distinct spectral regions. Namely, if $z$
is sufficiently far away from the singularities generated by
denominator $z - \epsilon - \Sigma^\parallel(z)|p|^2$ one can look
for the unknown $\gamma^{\alpha\beta}(z)$ in an approximate form,
replacing in (\ref{dP1}) the BEB CPA self-energy
$\Sigma^{\alpha\beta}(z)$ with its expression $\Sigma^{\alpha\beta
(0)}(z)$ for the disorder-free system. The solution found in this
way will be valid only for energies not very close to the band
edges of the disorder-free system. Around these points one can no
more neglect the disorder-induced corrections to the self-energy
component $\Sigma^{\parallel (0)} (z)$ in Eq.~(\ref{dP1}).
Nevertheless, the last situation is still addressable analytically
because near the band edges an explicit asymptotic relation
between $\bar \Gamma^{\alpha\beta}_{nn}(z)$ and
$\gamma^{\alpha\beta}(z)$ can be used.

\subsection{\label{sec:Iter} Iterative solution}

Away from the band edges of the disorder-free chain we can look
for the coherent-potential GT in the iterative form
\begin{equation}
\label{CPSepar} \gamma^{\alpha\beta}(z) =
\gamma^{(0)\alpha\beta}(z) + \gamma^{(1)\alpha\beta}(z).
\end{equation}
The first term with components
\begin{equation}
\label{Gamma0Def} \gamma^{(0)\parallel}(z) = \frac{|p|^2}{z -
\epsilon}, \qquad \gamma^{(0)\perp_i}(z) = 0,
\end{equation}
corresponds to the disorder-free system and
$\gamma^{(1)\alpha\beta}(z)$ is a small correction to be
determined. From equation (\ref{BarGSE}) linearized in the
parameters (\ref{ABCDef}) one has
\begin{equation}
\label{Gamma1}
\begin{array}{l}
\ds \gamma^{(1)\parallel}(z) = \frac{\bigl[z - \epsilon -
\Sigma^{(0)\parallel}(z)|p|^2\bigr]^2}{(z - \epsilon)^2} \left<
\tilde \Gamma^{(0)\parallel}_{nn}(z) \right>_2,
\\
\\
\ds \gamma^{(1)\perp_i}(z) = \left< \tilde
\Gamma^{(0)\perp_i}_{nn}(z) \right>_2,
\end{array}
\end{equation}
where $\tilde \Gamma^{(0)\alpha\beta}_{nn}(z)$ denotes the local
GT on a single impurity site in the disorder-free chain, related
to the disorder-free self-energy $\Sigma^{(0)\alpha\beta}(z)$ by
Eq.~(\ref{GSkobN}).

The computation of the disorder-free self-energy
$\Sigma^{(0)\alpha\beta}(z)$ is not straightforward due to the
fact that components $\gamma^{(0)\perp_i}(z)$ and $\bar
\Gamma^{(0)\perp_i}_{nn}(z)$ vanish leading to an undeterminate
expression in Eq.~(\ref{BarGSE}). To overcome this difficulty one
can employ the following arguments. Consider the disorder-free
chain with a single impurity molecule placed on the $n$th site in
such a way that its dipole moment $p^\alpha_n$ has no component
along the $\parallel$-axis. Since in this case the transfer energy
$J_{nm} = \sum_\alpha p^\alpha_n \vartheta^{\alpha\beta}_{nm}
p^\beta$ vanishes, the $n$th site, decoupled from the rest chain,
will be characterized by the local GT $\tilde
\Gamma^{(0)\alpha\beta}_{nn}(z)$ equal to the bare quantity
$p^\alpha_n g_n(z) p^\beta_n$. From Eq.~(\ref{GSkobN}), for any of
such $p^\alpha_n$, it follows that $\sum_\alpha p^\alpha_n
\Sigma^{(0)\alpha\beta}(z) p^\beta_n = 0$. Therefore, being a
diagonal tensor, the self-energy will have nonzero projection only
on the $\parallel$-axis, orthogonal to the subspace of the
considered $p^\alpha_n$. Using Eq.~(\ref{SigmaSimple}) for the
nontrivial component of the self-energy we end up with a simple
expression
\begin{equation}
\label{Sigma0} \Sigma^{(0)\parallel}(z) = \frac1{|p|^2} \! \left[z
- \epsilon - \frac1{\bar G^{(0)}_{nn}(z)}\right], \,\,
\Sigma^{(0)\perp_i}(z) = 0, \!\!
\end{equation}
where
\begin{equation}
\label{GFDF} \bar G^{(0)}_{nn}(z) = \int_{-\pi}^{\pi}
\frac{dk}{2\pi} \frac1{z - \epsilon - \vartheta_k^\parallel |p|^2}
\end{equation}
denotes the on-site scalar GF of the disorder-free chain. The
first of Eqs.~(\ref{Sigma0}) represents a natural relationship
between the scalar self-energy $z - \epsilon - 1/ \bar
G^{(0)}_{nn}(z)$ and the main component of the tensorial
self-energy in the absence of disorder.

Having derived the disorder-free self-energy we get the desired
correction (\ref{Gamma1}) to the coherent-potential GT in the form
\begin{equation}
\label{Gamma1Sol}
\begin{array}{l}
\begin{array}{l}
\ds \gamma^{(1)\parallel}(z) = \frac{a_2 |p|^2\bar
G^{(0)}_{nn}(z)}{(z - \epsilon)^2} + \frac{2 b_2^\parallel |p|}{(z
- \epsilon)^2}
\\
\ds \phantom{\gamma^{(1)\parallel}(z) =} + c_2^\parallel \left[ 4
\bar G^{(0)}_{nn}(z) - \frac3{z - \epsilon} \right],
\end{array}
\\
\\
\ds \gamma^{(1)\perp_i}(z) = c_2^{\perp_i} \bar G^{(0)}_{nn}(z).
\end{array}
\end{equation}
The final expression for the GT $\bar \Gamma^{\alpha\beta}_k(z)$,
valid in the low-disorder regime away from the band edges, is
obtained by inserting the total $\gamma^{\alpha\beta}(z)$ of
Eq.~(\ref{CPSepar}) into (\ref{CPA2Matrix}).

Let us comment on the analytical properties of the obtained
solution. The correction (\ref{Gamma1Sol}) obeys the large-$z$
asymptotics $\gamma^{(1)\alpha\beta}(z) \sim c^{\alpha\beta}_2 /
z$. Hence, the disorder-averaged GT behaves as $\bar
\Gamma^{\alpha\beta}_k(z) \sim (p^\alpha p^\beta +
c^{\alpha\beta}_2) / z$. Using definition (\ref{ABCDef}) for
$c^{\alpha\beta}_2$ it is easy to see that the coefficient
$p^\alpha p^\beta + c^{\alpha\beta}_2$ coincides with that of the
first term ($L^{\alpha\beta}_{k,0} = \left< p^\alpha_n p^\beta_n
\right>$) in the general high-energy expansion (\ref{LZAss}). That
is, even though for the zero-order solution the asymptotics $\bar
\Gamma^{(0)\alpha\beta}_k(z) \sim p^\alpha p^\beta / z$ is not
accurate, the correct form is restored as the first-order
iterative correction is taken into account. Furthermore, the next
two coefficients ($L^{\alpha\beta}_{k,1}$,
$L^{\alpha\beta}_{k,2}$) are also reproduced in the found solution
provided one neglects the third- and the higher-order centered
moments of the disorder distribution function in
Eqs.~(\ref{GamMom}). Nevertheless, the found GT can have in
general spurious poles outside the real axis around the band-edge
singularities of the disorder-free GF [Eq.~(\ref{GFDF})]. Such
nonphysical behavior is related to the fact that around the band
edge the second term in the expansion (\ref{CPSepar}) is no more
small as compared to the first. For the two components $\bar
\Gamma^{\perp_i}_k(z)$ the portion of spectral weight taken up by
the spurious poles vanishes upon decreasing the disorder strength
$c^{\perp_i}_2$. This guarantees the correct reproduction of the
optical absorption spectra for the corresponding polarizations, as
well as the DOS away from the band edges. The same is true for the
component $\bar \Gamma^\parallel_k(z)$ for a generic nonzero
momentum $k$. On the other hand, the spectral density component
$\bar A^\parallel_k(z)$ at momentum $k=0$ is itself strongly
concentrated near the energy
\begin{equation}
\label{ZetaDef} \zeta = \epsilon + \vartheta^\parallel_{k=0}
|p|^2,
\end{equation}
that usually represents an extremum of the bare exciton band.
Hence, the iterative solution found in this section does not
describe the most important part of the absorption profile for
polarization directed along the average dipole moment.

\subsection{\label{sec:Scaling} Scaling solution}

Let us now analyze the BEB-CPA equations near the lowest band edge
of the disorder-free system. We shall use the standard (for $J$,
but not $H$ aggregates) assumption that this point corresponds to
the center $k=0$ of the Brilloiun zone, thus coinciding with the
energy $\zeta$ defined in (\ref{ZetaDef}). Around $z = \zeta$ it
is convenient to represent the $\parallel$-component of the
coherent-potential GT as
\begin{equation}
\label{SigmaSmallIntr} \gamma^\parallel(z) = \frac{|p|^2}{z -
\epsilon - \sigma(z)}.
\end{equation}
Assuming the absolute value of the new unknown function
$\sigma(z)$ to be small compared to the total width of the exciton
band one can keep only a few significant terms in the expansion of
the coupling kernel $\vartheta^\parallel_k$ around $k=0$ while
performing the momentum integration in Eq.~(\ref{CPA2Matr}).
Provided the constant
\begin{equation}
\label{DeltaDef} \left. J = \frac{d^2 \vartheta^\parallel_k}{d
k^2} \right|_{k=0} |p|^2
\end{equation}
has a finite (positive) value, that is the case if the real-space
coupling $\vartheta^\parallel_{nm}$ decays faster than $|n -
m|^{-3}$ at large intermolecular separation, we can use the
effective mass approximation for the disorder-free exciton
dispersion
\begin{equation}
\label{EffMass} \vartheta^\parallel_k |p|^2 \approx
\vartheta^\parallel_{k=0} |p|^2 + \frac{J}2 k^2.
\end{equation}
This leads to the asymptotic expression for the corresponding
component of the local GT valid around the band edge:
\begin{equation}
\bar\Gamma^\parallel_{nn}(z) = \frac{|p|^2}{\sqrt{2 J} \sqrt{\zeta
- z + \sigma(z)} }.
\end{equation}
Furthermore, in accordance with Eq.~(\ref{SigmaSimple}), the
respective self-energy can be represented as
\begin{equation}
\Sigma^\parallel(z) = \frac1{|p|^2} \! \left[ z - \epsilon -
\sigma(z) - \sqrt{2 J} \sqrt{\zeta - z + \sigma(z) }\right]. \!
\end{equation}
Concerning the two components $\bar \Gamma^{\perp_i}_{nn}(z)$,
even though they share the singularity present in the above
$\bar\Gamma^\parallel_{nn}(z)$, one can demonstrate that it is
still possible to use the simple relationship
$\bar\Gamma^{\perp_i}_{nn}(z) = \gamma^{\perp_i}(z)$, or
equivalently to set $\Sigma^{\perp_i}(z) = 0$.

Writing the CPA self-consistency equation one can retain only the
most singular terms of Eq.~(\ref{dP1}) near the band edge $z
\approx \zeta$ where for vanishing disorder also $\sigma(z)
\approx 0$. The equation for $\gamma^{\perp_i}(z)$ reads simply
\begin{equation}
\label{NonImp} \gamma^{\perp_i}(z) =
\frac{c_2^{\perp_i}}{\sigma(z) + \sqrt{2 J} \sqrt{ \zeta - z +
\sigma(z)}}.
\end{equation}
As for the component $\gamma^\parallel(z)$, the only essential are
the first and the third terms in the right-hand side of
Eq.~(\ref{dP1}). By straightforward algebra we arrive at the
self-consistency condition
\begin{equation}
\label{MainScaling} \frac{\Delta^2 \, \sqrt{2 J} \sqrt{\zeta - z +
\sigma(z)}}{[\sigma(z) + \sqrt{2 J} \sqrt{\zeta - z +
\sigma(z)}]^2} = \sigma(z),
\end{equation}
that represents an equation to be solved for $\sigma(z)$. In this
formula we have used the shorthand notation
\begin{equation}
\label{DisorderCombined} \Delta = \sqrt{a_2 + 4 c_2^\parallel
\bigl( \vartheta^\parallel_{k=0} \bigr)^2 |p|^2}
\end{equation}
for a combination which can be thought of as an effective disorder
strength parameter governing the low-disorder limit of BEB CPA
near the band edge.

The algebraic structure of equation (\ref{MainScaling}) is still
too complicated to obtain its solution in a closed form. At the
same time the scaling of $\sigma(z)$ with the disorder strength
$\Delta$ can be found easily. Namely, in the interval $\bigl|z -
\zeta \bigr| \lesssim J (\Delta / J)^{4/3}$ the solution (both
real and imaginary part) behaves as
\begin{equation}
\label{ScalingSol} \sigma(z) \sim J \left( \frac{\Delta}{J}
\right)^{4/3}.
\end{equation}
Employing this scaling relation one can find the low-disorder
asymptotic expressions for the absorption linewidth and the
exciton coherence length in the resonance region. Substituting
(\ref{SigmaSmallIntr}) into Eq.~(\ref{CPA2Matrix}) the main
component of the disorder-averaged GT around the phase-space point
$z \approx \zeta$, $k \approx 0$ can be written as
\begin{equation}
\label{GammaScalReg} \bar \Gamma^\parallel_k(z) = \frac{|p|^2}{z -
\zeta - \sigma(z) - J k^2/2}.
\end{equation}
The corresponding polarizability
\begin{equation}
\bar\chi^\parallel(z) = -{\cal N}\frac{|p|^2}{z - \zeta -
\sigma(z)},
\end{equation}
obtained from the basic relation (\ref{DAChiFour}), is
characterized by a resonant behavior near the band edge. The width
of the resonance, estimated as $\eta^\parallel \sim -\Im
\sigma(\zeta)$, scales with the disorder strength $\Delta$ as
\begin{equation}
\label{EtaScal} \eta^\parallel \sim J \left( \frac{\Delta}{J}
\right)^{4/3}.
\end{equation}
Furthermore, one can make use of the effective mass approximation
(\ref{EffMass}) to get a scaling of the exciton coherence length
in the spectral region around the absorption resonance. According
to the arguments of Sec.~\ref{sec:Anal} applied to expression
(\ref{GammaScalReg}), the complex wave number $k=\xi^\parallel(z)$
governing the asymptotic behavior of the disorder-averaged GT at
large intermolecular separation is found from equation
\begin{equation}
z - \zeta - \sigma(z) - \frac{J}2 k^2 = 0.
\end{equation}
Taking $z \approx \zeta$ we end up with the following scaling of
the coherence length in the resonance region:
\begin{equation}
\label{NScal} N^\parallel \sim \left( \frac{\Delta}{J}
\right)^{-2/3}.
\end{equation}
It should be also mentioned that combining Eq.~(\ref{EtaScal})
with (\ref{NScal}) leads to the scaling relation
\begin{equation}
\label{NETAScal} \eta^\parallel \sim J \bigl( N^\parallel
\bigr)^{-2},
\end{equation}
which does not involve the disorder strength $\Delta$. This
dependence is universal in the sense that it follows only from the
shape of the exciton dispersion in the disorder-free system in the
vicinity of the band edge. Indeed, the scaling (\ref{NETAScal})
can be established without rigorous computation of the spectral
density but by simply noticing that $\eta^\parallel$ and
$1/N^\parallel$ can be considered, respectively, as uncertainties
of the exciton energy and momentum brought about by the disorder
scattering. The relation between these quantities imposed by the
dispersion (\ref{EffMass}) is of the quadratic form
(\ref{NETAScal}).

It can be noticed that the terms of expansion (\ref{LDMain})
proportional to the higher-order centered moments of the disorder
distribution will bring stronger singularities in the vicinities
of the band edges compared to ones of the second-order centered
moments. Nevertheless, the dominant contribution to the solution
will come solely from the second-order moments. Let us illustrate
this for the simple situation when only diagonal disorder is
present. A derivation analogous to that leading to
Eq.~(\ref{ScalingSol}) shows that the centered moment $a_\nu$ of
order $\nu > 2$ alone would produce the scaling $\sigma_\nu(z)
\sim J (a_\nu / J^\nu)^{2/(1+\nu)}$. If the probability to
encounter a random energy $\epsilon_n$ falls off rapidly enough
away from the mean value $\epsilon$ we have $a_\nu \sim
a_2^{\nu/2}$. Thus, the contribution $\sigma_\nu(z) \sim J (a_2 /
J^2)^{\nu/(1+\nu)}$ will be always suppressed by that of the
second-order moment $\sigma(z) \sim J (a_2 / J^2)^{2/3}$
considered so far.

When applied to a molecular chain with only diagonal disorder the
asymptotic expressions (\ref{EtaScal}) and (\ref{NScal}) coincide
with those obtained in Ref.~\onlinecite{MalyshevScal} using
different scaling arguments. Furthermore, for the case of diagonal
disorder the asymptotics (\ref{EtaScal}) has been found in
Ref.~\onlinecite{Huber2} by means of the scalar CPA. Our theory
thus provides an extension of these results treating both the
on-site energy disorder and the disorder in transition dipoles on
an equal footing through the combined parameter $\Delta$ of
Eq.~(\ref{DisorderCombined}). The fact that BEB CPA is able to
reproduce the known scaling relations is not surprising because
this self-consistent approximation becomes exact in the
disorder-free limit.

\section{\label{sec:Results} Illustrative applications}

\subsection{\label{sec:SimpleModel} Model and simplified parametrization}

\begin{figure}
\centering
\includegraphics[width = 5.0cm]{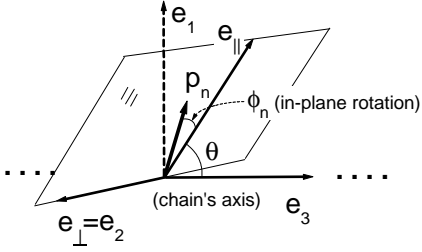} \caption{\label{Fig2}
The studied geometry of orientationally disordered molecular
chain. The dipoles are distributed in the plane built on vectors
$e_\parallel^\alpha$, $e_\perp^\alpha$ and forming the angle
$\theta$ with the chain's direction $e_3^\alpha$.}
\end{figure}

In this section we demonstrate how the general theory discussed so
far can be applied to a specific model of disordered aggregate: a
chain with purely orientational disorder in the transition
dipoles. In addition to having constant absolute values the random
dipoles will be assumed to lie in parallel planes forming an angle
$\theta$ with the chain's direction [Fig.~\ref{Fig2}]. The random
in-plane orientation angles $\phi_n$ will be modelled by a box
probability with density $1/(2 \Phi)$ in the interval $|\phi_n| <
\Phi$ and zero outside. We shall consider the molecules to
interact as dipoles in an isotropic background, so that the
tensorial kernel $\vartheta^{\alpha\beta}_{nm}$ acquires the form
\begin{equation}
\label{DDKernel}
\begin{array}{c}
\ds \vartheta^{11}_{nm} = \vartheta^{22}_{nm} = V_{nm}, \quad
\vartheta^{33}_{nm} = -2 V_{nm},
\\
\ds \vartheta^{12}_{nm} = \vartheta^{13}_{nm} =
\vartheta^{23}_{nm} = 0. \phantom{\int\limits^\pi \!\!\!\!}
\end{array}
\end{equation}
For the long-range (LR) dipole-dipole coupling one would have
$V_{nm} = |n-m|^{-3}$. However, since in this case the momentum
integral (\ref{CPA2Matr}) can not be reduced to a closed analytic
form, we shall restrict ourselves to the more easily treatable
nearest-neighbor (NN) [$V_{nm} = \delta_{|n-m|,1}$] or
next-nearest-neighbor (NNN) [$V_{nm} = \delta_{|n-m|,1} +
\delta_{|n-m|,2} / 8$] couplings. For concreteness we shall always
assume that $\cos^2\theta > 1/3$, so that the disorder-free chain
($\Phi = 0$) represents a $J$ aggregate. The singular case $\cos^2
\theta = 1/3$ is considered separately in Appendix \ref{app:Sing}.
From now on the energy will be measured in the units of the NN
interaction strength, and the dipoles will be chosen to have unit
absolute values. We also shift the reference point of energy to
the position of the bare molecular level.

Since the random dipoles are restricted in a single plane the
theory can be conveniently reformulated in terms of tensors
projected onto the corresponding two-dimensional subspace. We
choose the new basis as
\begin{equation}
\begin{array}{c}
\ds e^1_\parallel = \sin\theta, \qquad e^2_\parallel = 0, \qquad
e^3_\parallel = \cos\theta,
\\
\ds e^1_\perp = 0, \qquad e^2_\perp = 1, \qquad e^3_\perp = 0,
\phantom{\int\limits^\pi \!\!\!\!}
\end{array}
\end{equation}
i.e., the first vector is parallel to dipoles of the disorder-free
chain, the second is orthogonal to it and to the chain's axis. In
terms of the projected components $p^\parallel_n = \cos\phi_n$,
$p^\perp_n = \sin\phi_n$ the three-dimensional dipoles are given
by
\begin{equation}
\label{DMom} p^\alpha_n = e^\alpha_\parallel p^\parallel_n +
e^\alpha_\perp p^\perp_n.
\end{equation}
The advantages of this representation come from the fact that our
disorder distribution density is invariant under the change of
sign of $p^\perp_n$. Hence, since upon projection the coupling
(\ref{DDKernel}) remains diagonal,
\begin{equation}
\label{DDKernelPrime} \vartheta^\parallel_{nm} = (1 - 3
\cos^2\theta) V_{nm}, \qquad \vartheta^\perp_{nm} = V_{nm},
\end{equation}
the reduced BEB CPA will involve only diagonal tensors. As a
result we end up with the affordable equations, whose explicit
form for the case of NN and NNN coupling is given in Appendix
\ref{BEBExpl}. Moreover, as the additional symmetry requirements
formulated in Sec.~\ref{sec:WeakDis} are fulfilled, the
low-disorder solution derived there can be employed. Since all
components of a physically-measurable tensor can be
straightforwardly reconstructed using Eq.~(\ref{DMom}) from the
two diagonal components of the corresponding projected tensor, the
last two will be presented every time to illustrate the outcome of
the theory.

\subsection{\label{sec:Numer} Numerical results}

Due to their explicit form the equations of Appendix \ref{BEBExpl}
can be efficiently solved using standard numerical methods. In
order to test the accuracy of the theory we also performed exact
diagonalization for an open chain of ${\cal N} = 250$ molecules
with either NN or LR coupling. The statistical error was
reasonably small after averaging over 5000 disorder realizations
with the energy domain divided into 400 output intervals.

\begin{figure}
\centering
\includegraphics[width=8cm]{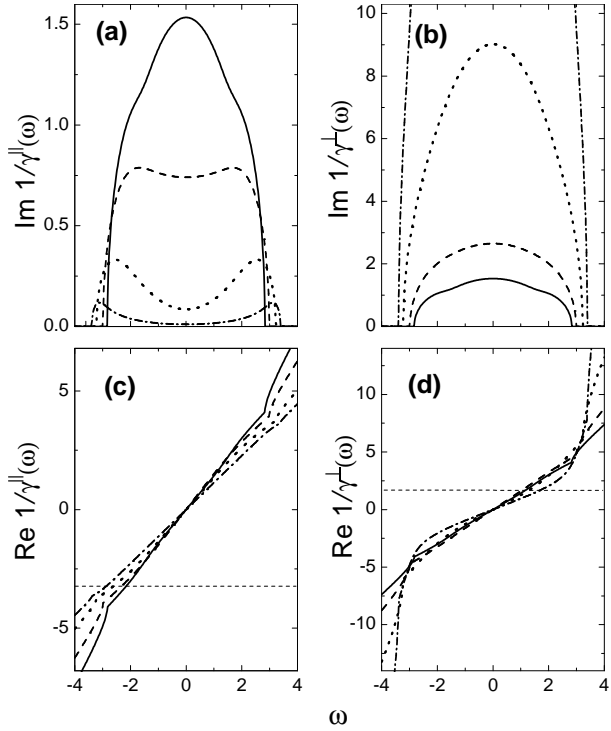}
\caption{\label{fig:Self} The imaginary (a,b), and real (c,d)
parts of the BEB-CPA self-energy for the case of NN coupling.
Several values of the disorder parameter are used: $\Phi = 0.6$
(solid lines), $\Phi = 0.9$ (dashed lines), $\Phi = 1.3$ (dotted
lines), $\Phi = 3.14$ (dash-dotted lines). $\theta = 0.26$ for all
curves. The horizontal lines in panels (c) and (d) mark the
diagonal elements of the NNN coupling matrix at momentum $k=0$:
$\vartheta^\parallel_{k=0} = -3.603$, $\vartheta^\perp_{k=0} =
2$.}
\end{figure}

The tensorial components of the coherent-potential GT are plotted
against energy in Fig.~\ref{fig:Self}. This figure can be used to
illustrate analyticity of the theory and to predict certain
features of the physical quantities to be presented below.
Specifically, the functions $\Im 1/\gamma^\parallel(\omega)$ and
$\Im 1/\gamma^\perp(\omega)$ obey the correct sign that guarantees
the positivity of DOS and optical absorption. From
Eq.~(\ref{CPA2Matrix}) it follows that the resonant denominators
of the exciton spectral density components are given by $[ \Re
1/\gamma(\omega) - \vartheta_k]^2 + [ \Im 1/\gamma(\omega) ]^2$,
where polarization labels $\parallel$ or $\perp$ are assumed.
Hence, provided $\Im 1/\gamma(\omega) \ll 1$, the spectral density
will be characterized by a sharp resonance located around the
frequency $\omega$ to be found from $\Re 1/\gamma(\omega) -
\vartheta_k = 0$. The solution of this equation for either
polarization and momentum $k$ can be found graphically from
Fig.~\ref{fig:Self}(c,d) [where, in particular, the positions of
$\vartheta^{\parallel}_k$ and $\vartheta^{\perp}_k$ at $k=0$ are
shown with horizontal lines]. Actually, the assumption on
smallness of the imaginary part of the inverse coherent-potential
GT is valid only in the case of component
$\gamma^\parallel(\omega)$ and only for moderately weak disorder.
As concerns $\gamma^\perp(\omega)$, the above condition is never
satisfied, so that one ends up with no well-defined quasiparticle
peak in the corresponding component of the spectral density.

\begin{figure}
\centering
\includegraphics[width=8cm]{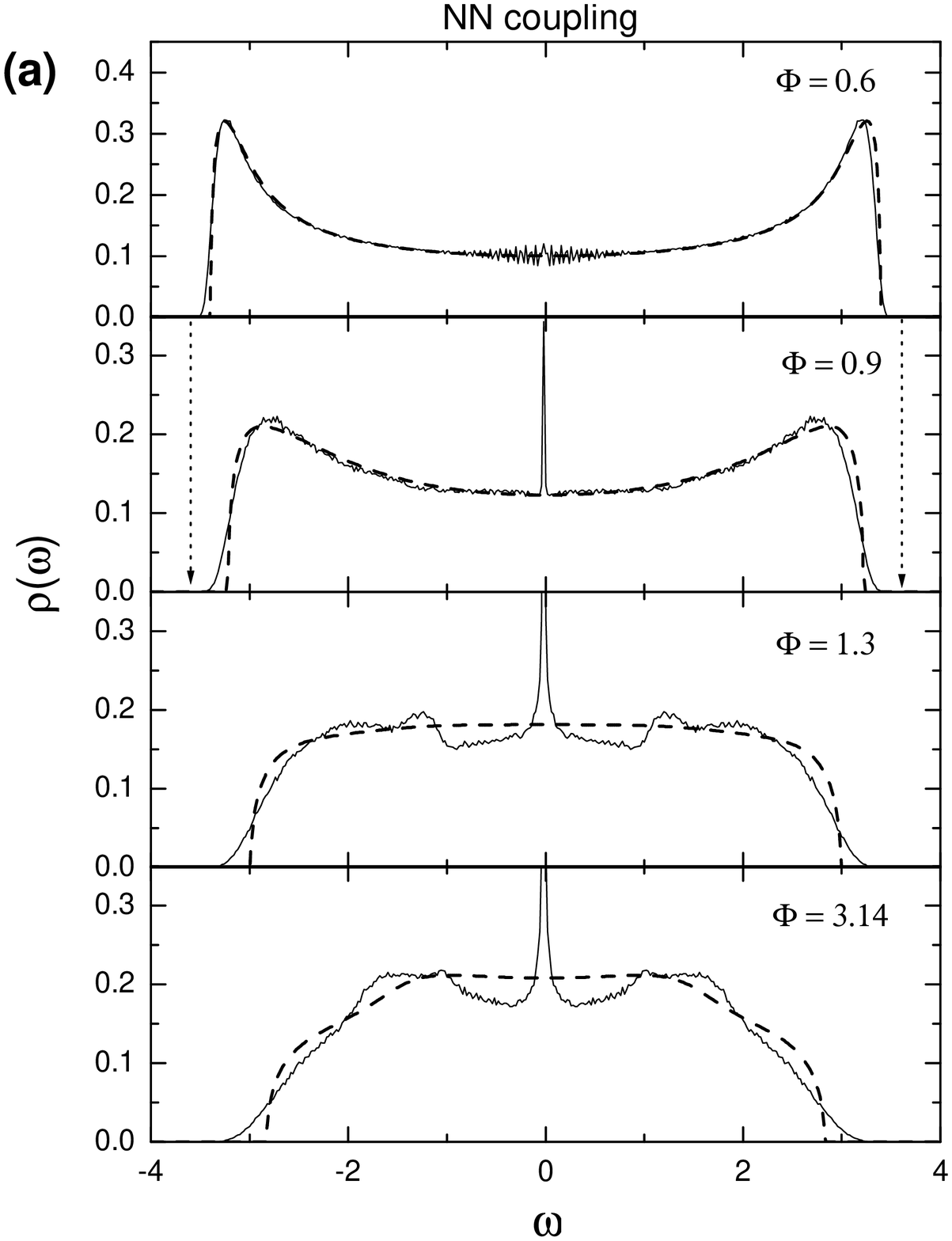}
\includegraphics[width=8cm]{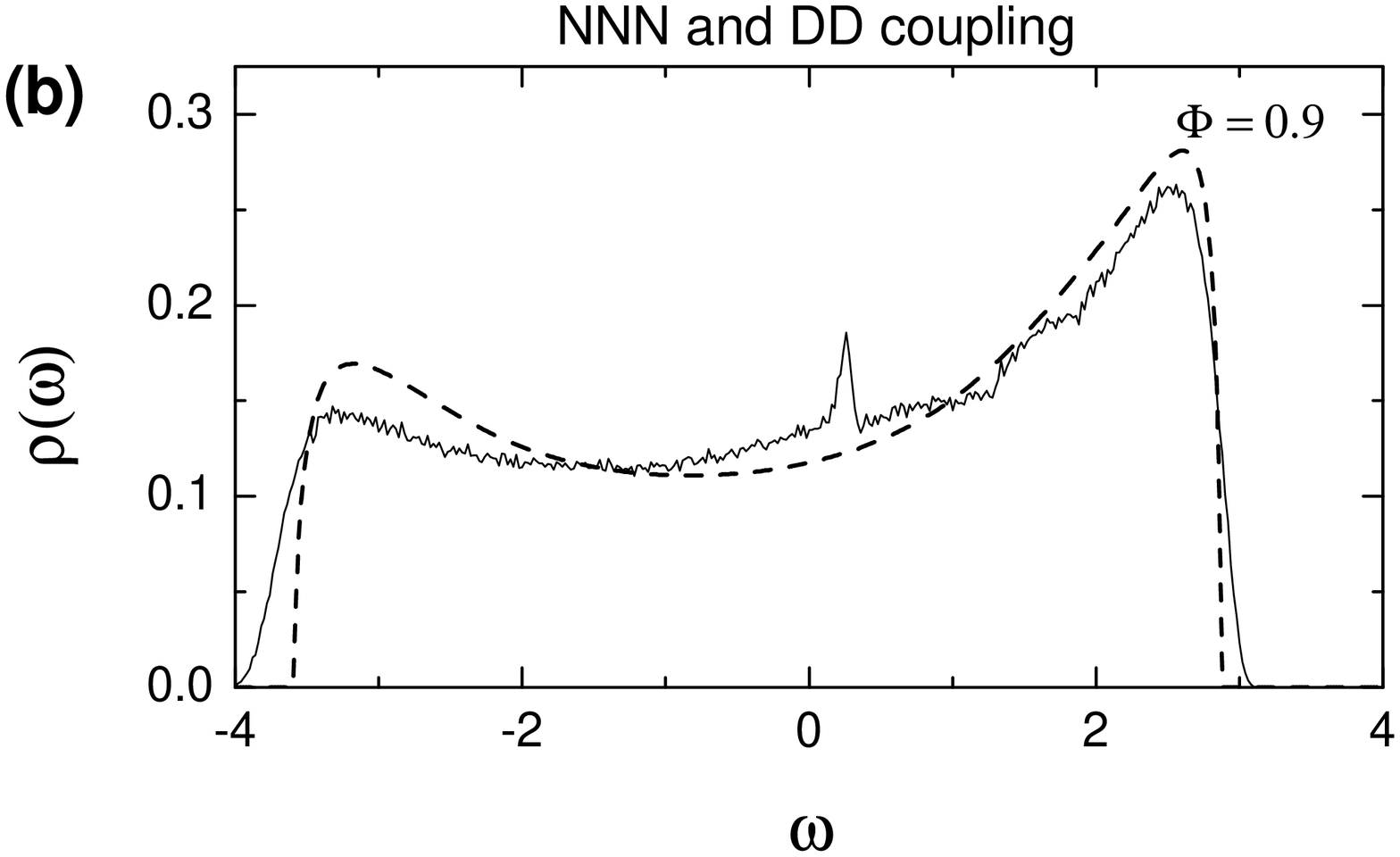}
\caption{\label{fig:Rho} The disorder-averaged exciton DOS for
several values of disorder parameter $\Phi$ calculated with BEB
CPA (dashed lines) or exact diagonalization (solid lines). (a)
Case of NN coupling (both BEB CPA and exact diagonalization). (b)
Case of NNN coupling (BEB CPA) or LR coupling (exact
diagonalization). $\theta = 0.26$ for all curves. The arrows mark
the spectral bound estimates for the case of NN coupling and $\Phi
= 0.9$, namely, $\omega = \pm 3.60$.}
\end{figure}

The profiles of the disorder-averaged exciton DOS are presented in
Fig.~\ref{fig:Rho}. Except for the well-known band-center Dyson
singularity\cite{Dyson} in the case of NN coupling and a similar
but nonsingular feature for the LR coupling\cite{KozlovRef} the
numerical DOS is reproduced well within BEB CPA. As a confirmation
of the analyticity of the theory no unphysical behavior of the DOS
has been observed. The theory also captures properly the exact
symmetry of the spectrum present in any tight-binding system
without diagonal disorder and with only NN coupling. The
asymmetric DOS in the case of LR interactions is satisfactorily
approximated already by that of the NNN coupling. Furthermore, it
can be observed that the upper and lower estimates of the spectrum
region mentioned in Sec.~\ref{sec:Anal} are not violated in the
presented solution. For instance, in the case of NN coupling the
spectrum should not extend beyond the interval $|\omega| < 2 \max
|J_{n \, n+1}|$. The maximal absolute value of coupling accessible
by varying parameters $\phi_n$ within the interval $|\phi_n| <
\Phi$ is given differently in the distinct domains of the
parametric space: if $\cos^2\theta > 2/3$ it equals to $3
\cos^2\theta - 1$, while in the opposite case it becomes $|1 - 3
\cos^2\theta| \cos^2 \Phi + \sin^2\Phi$ for $\Phi \le \pi/2$, or
$1$ for $\Phi > \pi / 2$.

\begin{figure}
\centering
\includegraphics[width=8cm]{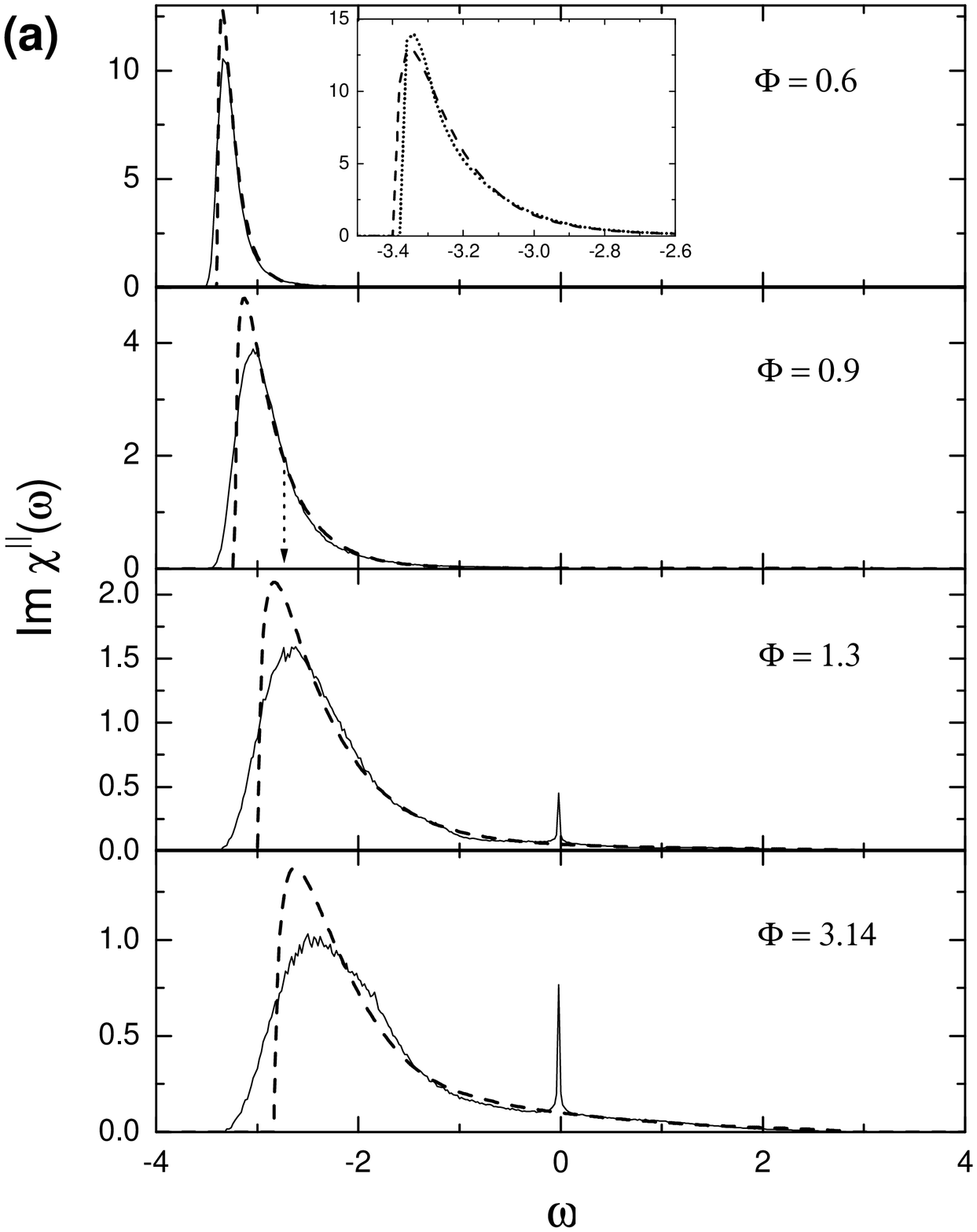}
\includegraphics[width=8cm]{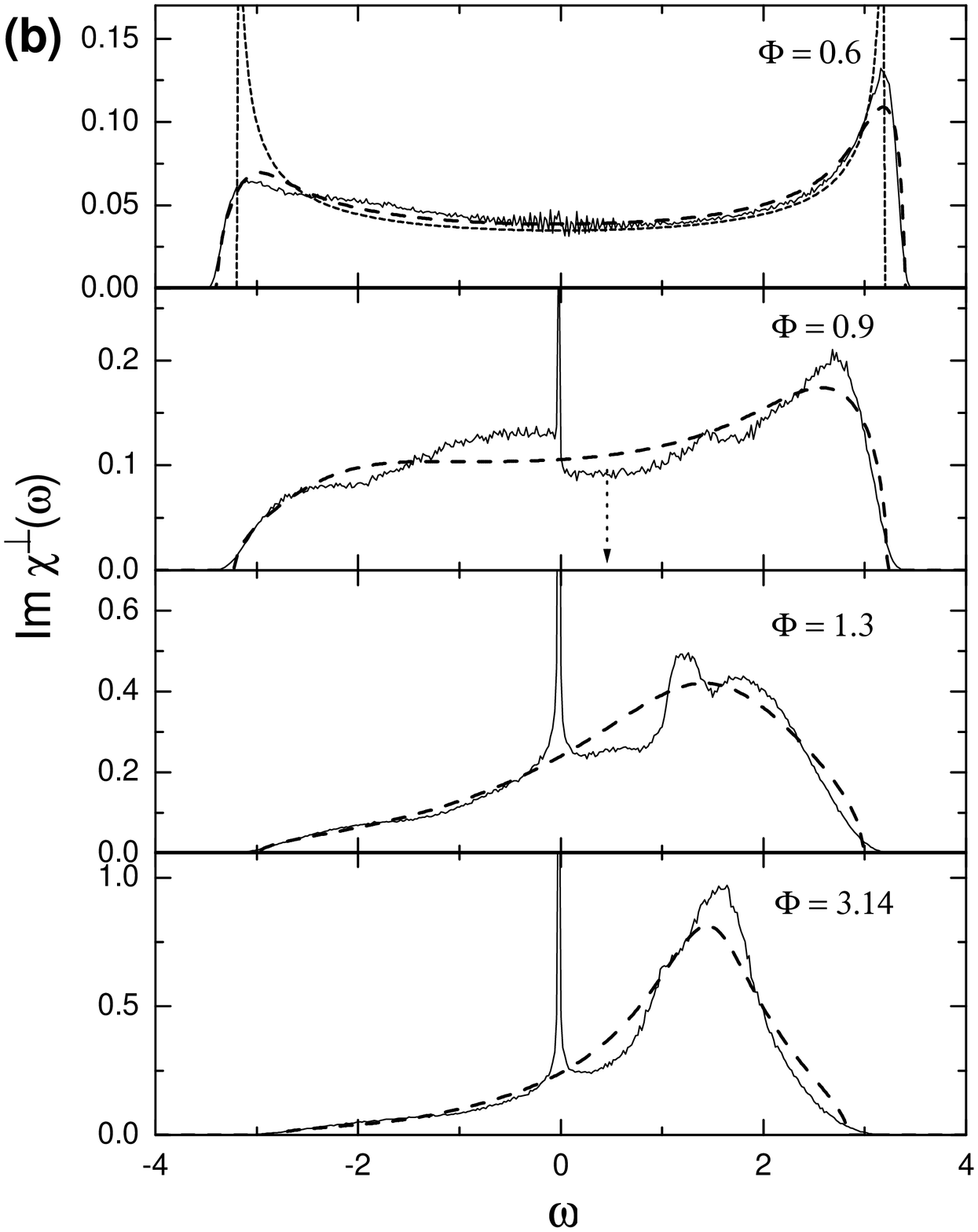}
\caption{\label{fig:Chi} The excitonic absorption spectra for NN
coupling and several values of disorder parameter $\Phi$
calculated with BEB CPA (dashed lines) or exact diagonalization
(solid lines). $\theta = 0.26$ for all curves. The arrows mark
estimates of the spectral center for $\Phi=0.9$: $\Omega^\parallel
= -2.77$, $\Omega^\perp = 0.46$. For $\Phi = 0.6$ the absorption
spectra resulting from the low-disorder limit of BEB CPA are
presented (dotted lines).}
\end{figure}

The disorder-averaged absorption spectra (normalized by the number
of molecules in the chain) along the two essential polarizations,
$e^\alpha_\parallel$ and $e^\alpha_\perp$, are presented in
Fig.~\ref{fig:Chi}. The component $\Im
\bar\chi^\parallel(\omega)$, corresponding to the direction of
preferable orientation of the dipoles, demonstrates behavior
typical for disordered $J$ aggregates: a sharp resonance acquiring
inhomogeneous width upon increasing disorder is located near the
bottom of the excitonic band. As for the polarization orthogonal
to the preferred orientation of the dipoles, the absorption
component $\Im \bar\chi^\perp(\omega)$ is generated solely by
orientational disorder. In contrast to the previous case no narrow
resonance is observed, but rather a broad spectrum shifted to
energies higher with respect to the bare molecular level. The
positions of the absorption maxima can be roughly estimated from
the zero and first moments of the spectral density as
$\Omega^\parallel \sim L^\parallel_{k=0,1} / L^\parallel_{k=0,0}$,
$\Omega^\perp \sim L^\perp_{k=0,1} / L^\perp_{k=0,0}$. Using
Eqs.~(\ref{Moms}) for these parameters in the simplest case of NN
coupling we end up with $\Omega^\parallel = (1-3\cos^2\theta)[1 +
\sin2\Phi / (2\Phi)]$, $\Omega^\perp = 1 - \sin2\Phi / (2\Phi)$.
These expressions one more time confirm that the centra of the
absorption spectra are located at essentially different energies
for the two polarizations.

\begin{figure*}
\centering $\phantom{xxxx}$ \includegraphics[width=6cm]{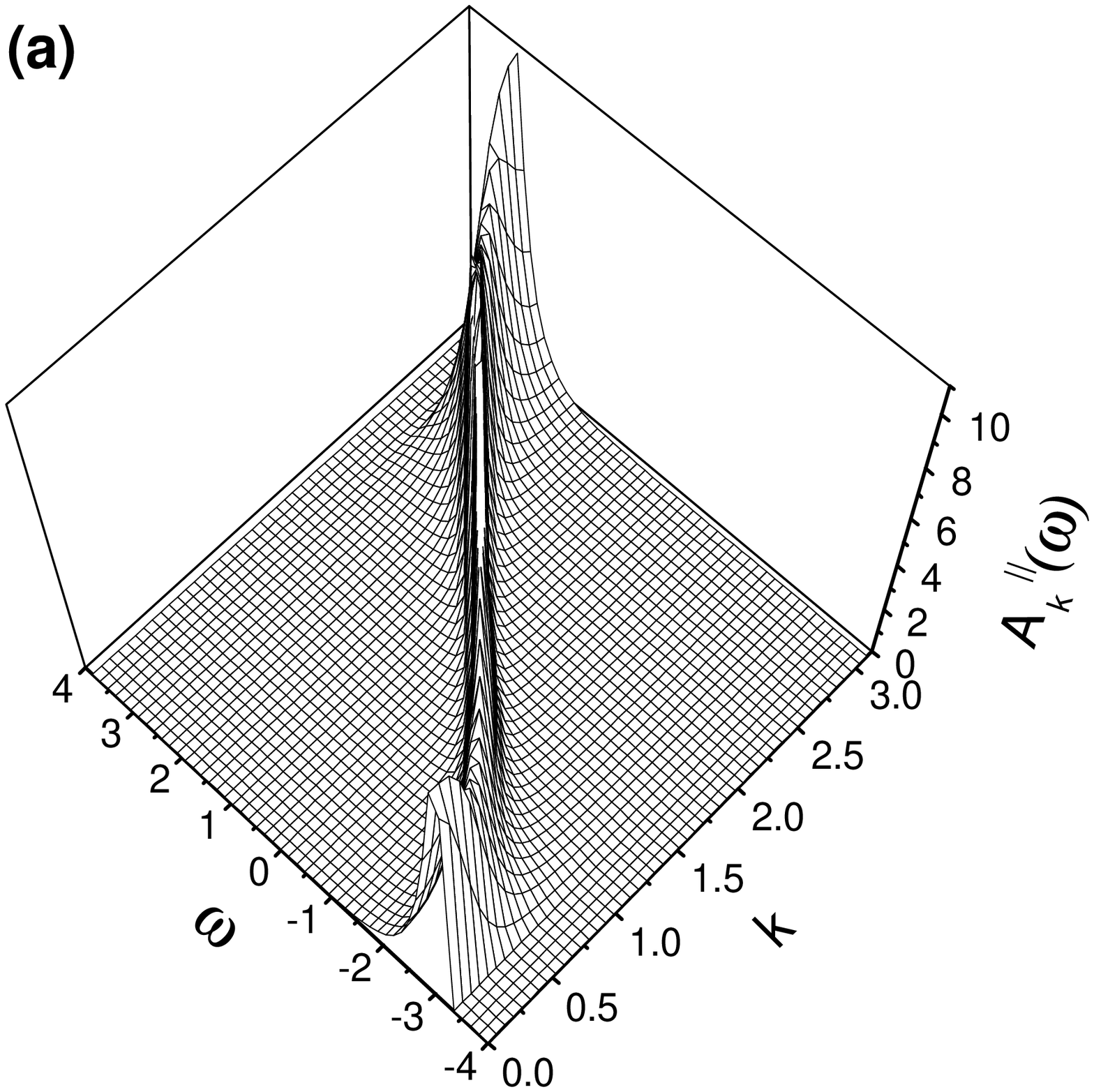}
$\phantom{xxxxxxx}$ \includegraphics[width=6cm]{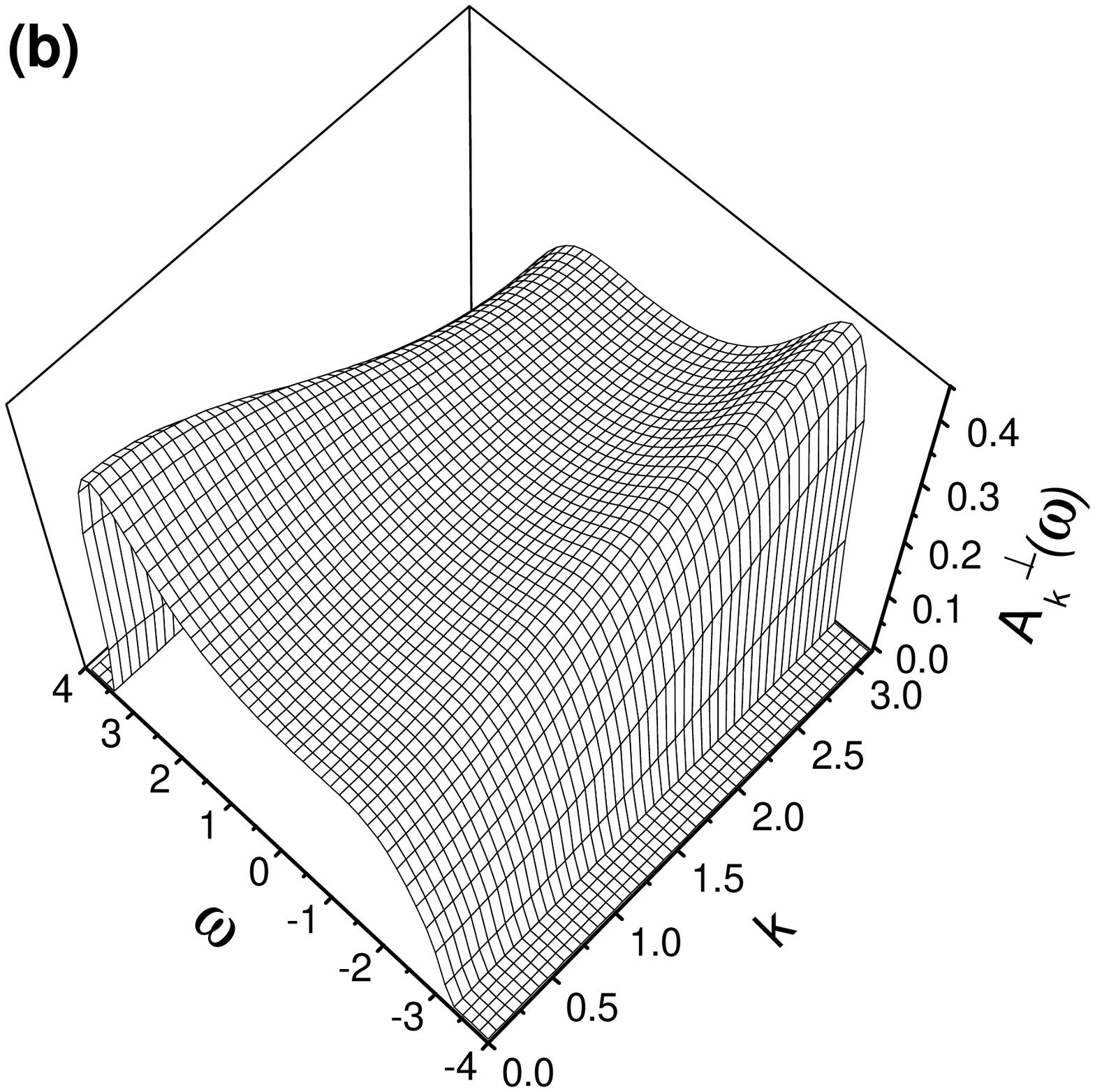}
\includegraphics[width=6.5cm]{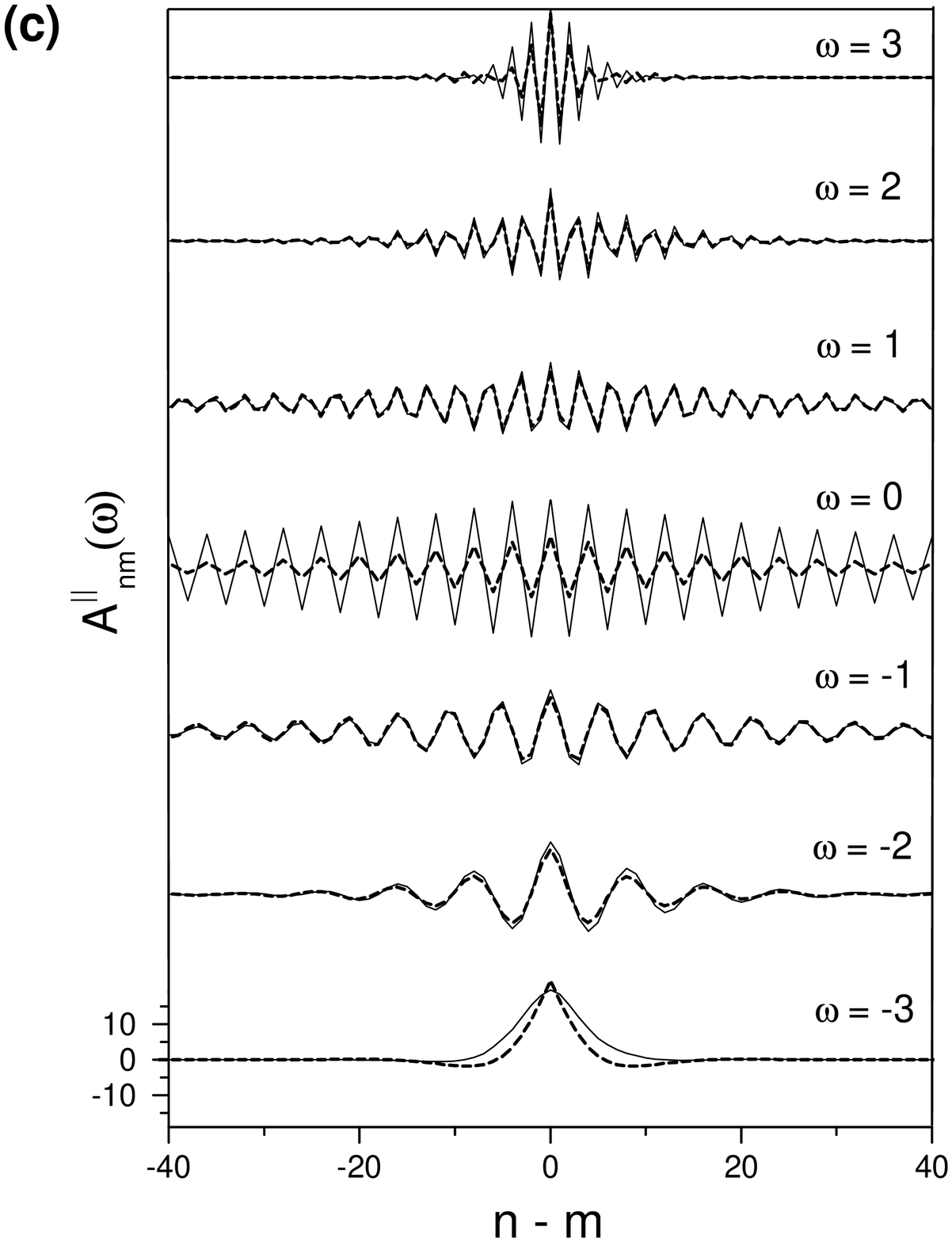} $\phantom{xxxxxx}$
\includegraphics[width=6.5cm]{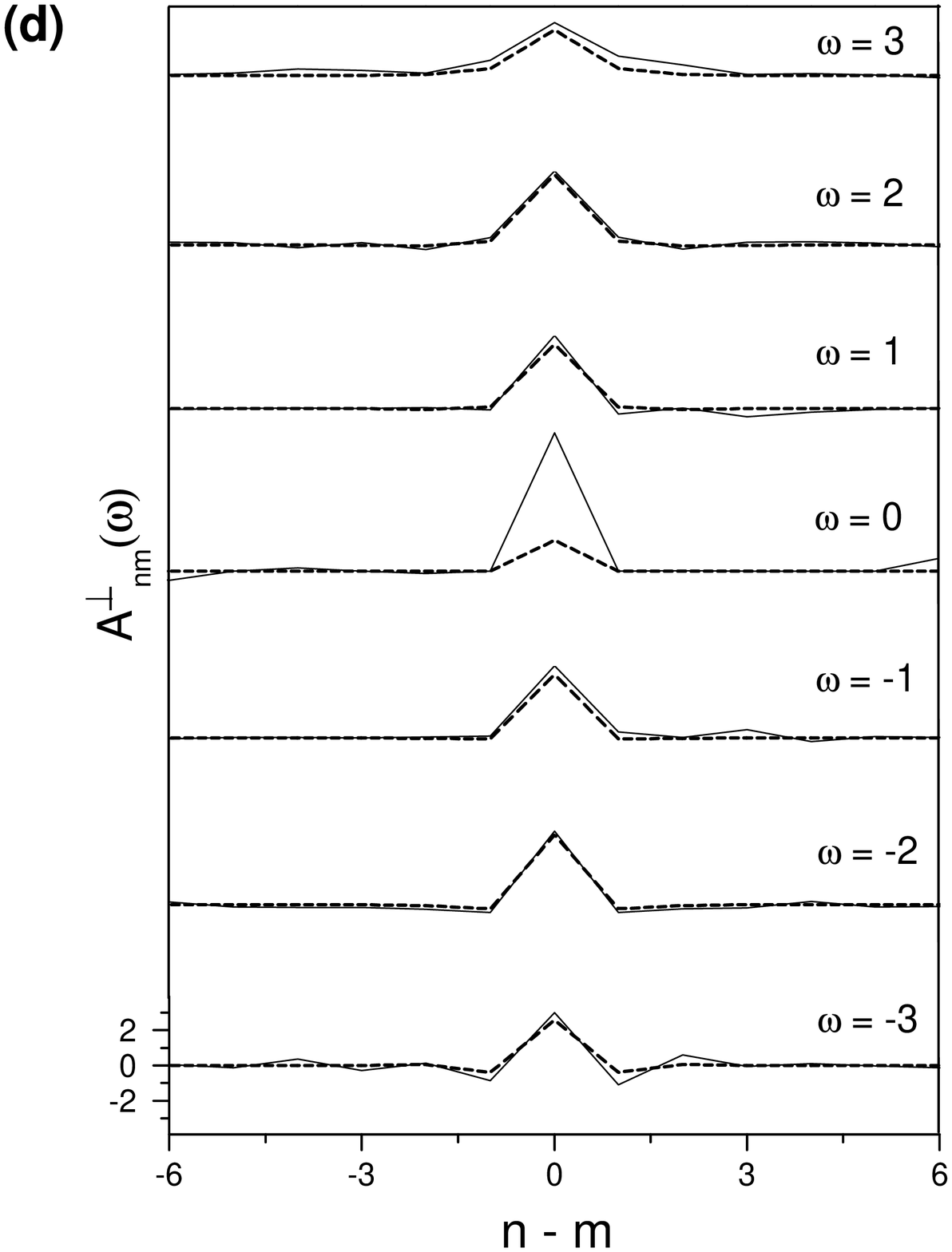} $\phantom{xxxxx}$
\caption{\label{fig:GF} (a,b) The tensorial components of the
BEB-CPA exciton spectral density for the case of NN coupling. Only
positive momenta $k$ are shown. (c,d) The disorder-averaged
spectral density in the real-space representation for a few fixed
energies $\omega$. The BEB-CPA results (dashed lines) are
confronted with ones of the exact diagonalization (solid lines).
The parameters are $\Phi = 0.9$, $\theta = 0.26$. [The strong
discrepancy at $\omega = 0$ is the effect of the Dyson singularity
not reproduced by the BEB CPA.]}
\end{figure*}

To provide additional understanding of the features observed in
the absorption spectra, the complete BEB-CPA spectral density is
plotted in Fig.~\ref{fig:GF}(a,b). For the considered disorder
parameter ($\Phi = 0.9$) the component $\bar
A_k^\parallel(\omega)$ is characterized by a well-pronounced
quasiparticle structure with most of the spectral wight located
around the disorder-free exciton branch $\omega =
\vartheta^\parallel_k$. As for the part $\bar A^\perp_k(\omega)$,
similarly to the already considered case of momentum $k=0$ (the
only directly accessible in the linear optics), the function has
no resonant behavior in all the phase space. The spectral density
transformed to the real-space representation is shown in
Fig.~\ref{fig:GF}(c,d).  The component $\bar
A^\parallel_{nm}(\omega)$ is essentially nonzero in the interval
$|n-m| \lesssim 100$, and for fixed exciton energy has oscillatory
behavior as a function of intermolecular distance $n - m$.
Starting from zero near the bottom of the band the number of nodes
monotonically increases as one moves towards the upper band edge.
Since absorption is obtained from the real-space spectral density
after the summation (\ref{DAChi}), the absence of oscillations
explains the strong absorbance for the $||$-polarization near the
band bottom. As concerns polarization orthogonal to the preferred
orientation of the transition dipoles, the function $\bar
A^\perp_{nm}(\omega)$ is confined on a few ($|n - m| \lesssim 1
\div 2$) sites. This implies the absence of the coherent optical
response of the molecules resulting to a weak and very broad
absorption for this polarization direction.

\begin{figure}
\centering
\includegraphics[width=8.5cm]{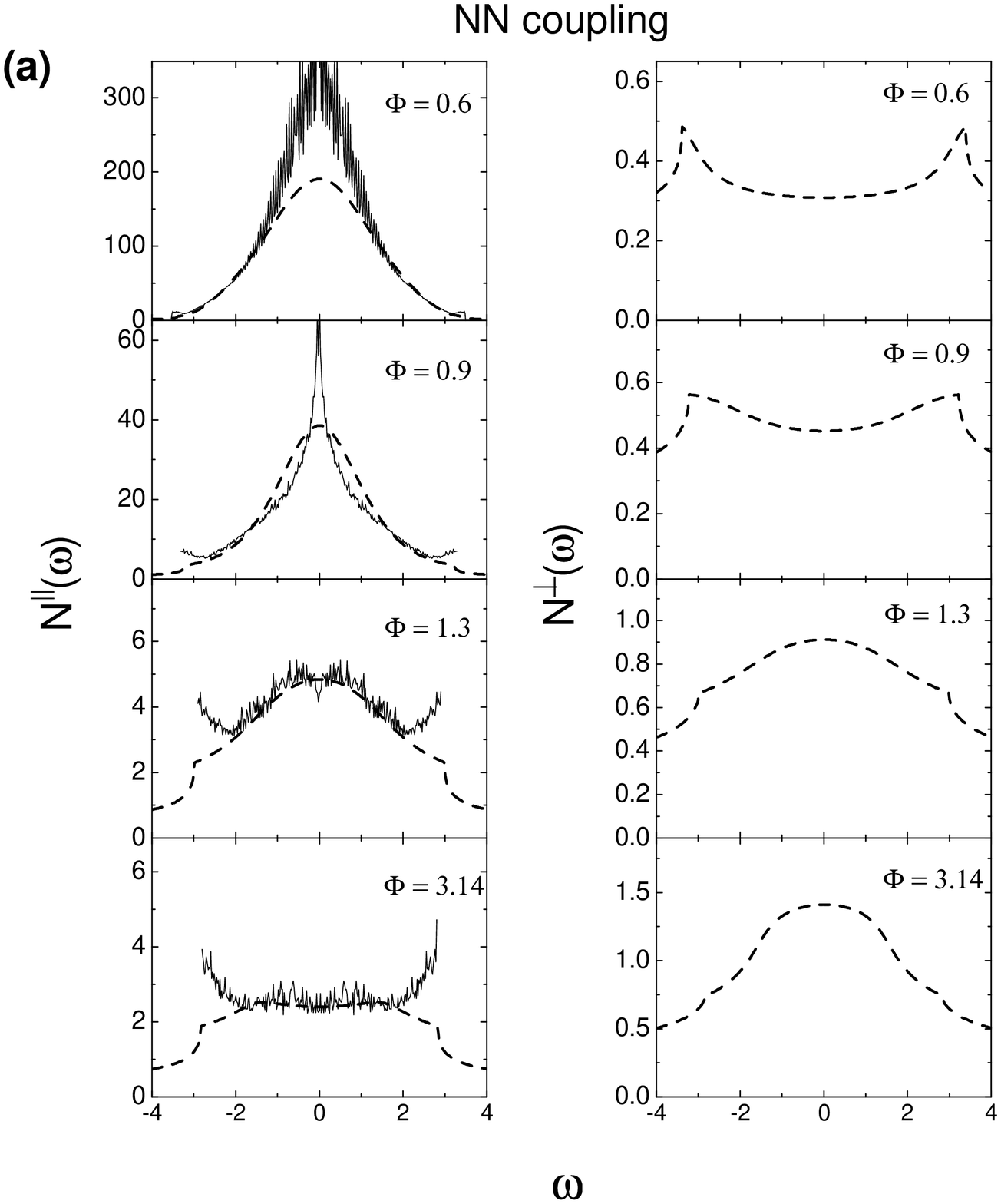}
\includegraphics[width=8.5cm]{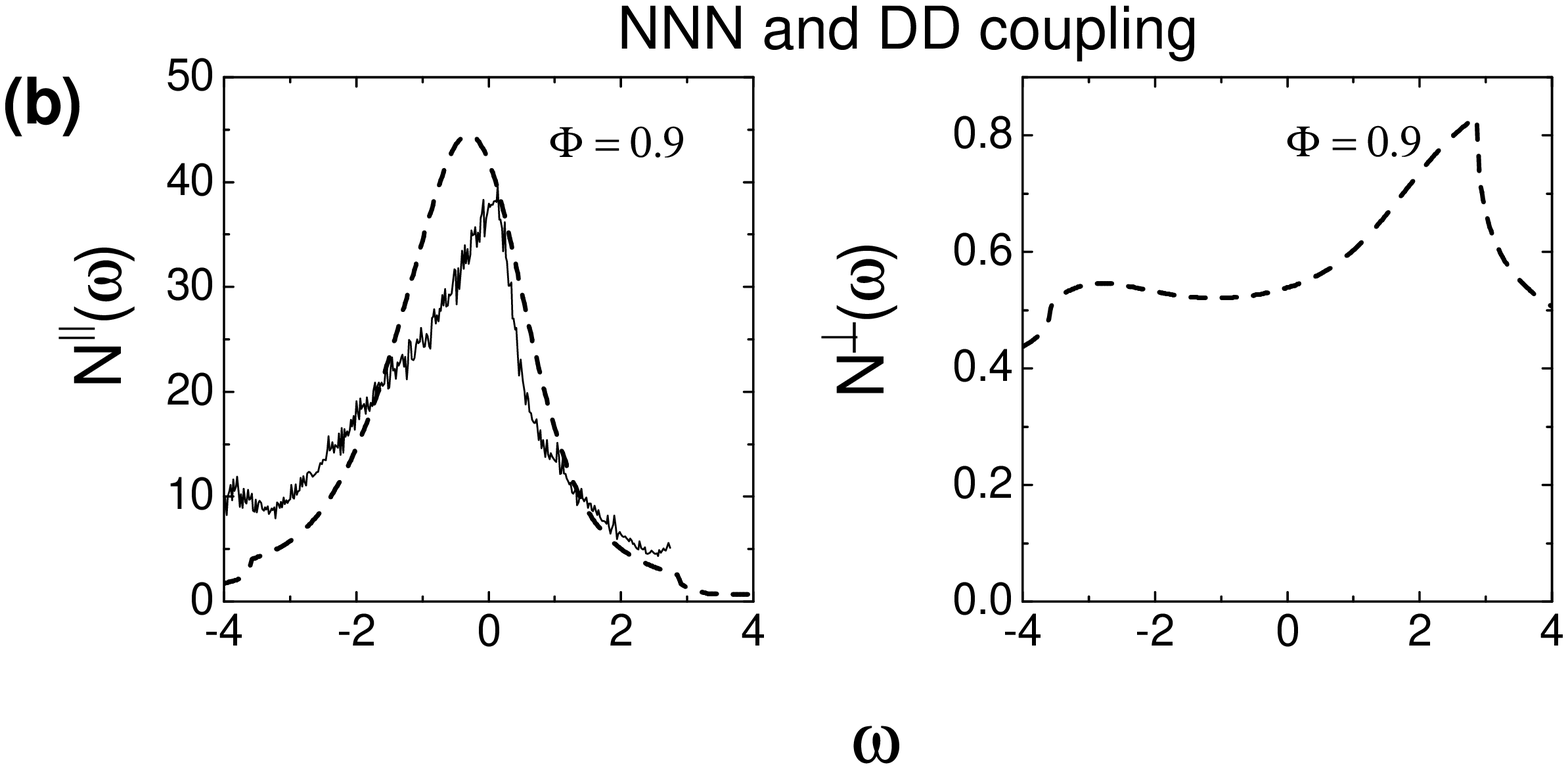}
\caption{\label{fig:Coh} Energy dependence of the exciton
coherence length for several values of the disorder parameter
$\Phi$. The BEB-CPA results are presented in the case of both
$N^\parallel$ and $N^\perp$ (dashed lines) and ones of the exact
diagonalization (solid lines) only for $N^\parallel$. $\theta =
0.26$ for all curves.}
\end{figure}

Figure \ref{fig:Coh} illustrates variation of the exciton
coherence length across the spectrum. The BEB-CPA coherence length
has been found extracting the parameters $\xi^{\alpha\beta}(z)$ of
the large-intersite-separation asymptotics (\ref{NCohDef}) from
Eq.~(\ref{CharEq}). Owing to the symmetry of the problem the
latter splits into the two independent equations $1 -
\gamma^\parallel(z) \vartheta^\parallel_k = 0$, $1 -
\gamma^\perp(z) \vartheta^\perp_k = 0$. Consequently, each of the
two components $\bar \Gamma^\parallel_{nm}(z)$ and $\bar
\Gamma^\perp_{nm}(z)$ is characterized by its own coherence
length, to be denoted respectively as $N^\parallel(\omega)$ and
$N^\perp(\omega)$. From the results presented in
Fig.~\ref{fig:Coh} it follows that $N^\parallel(\omega)$ strongly
depends on energy, having maximum around the band center and
decreasing as $\omega$ approaches the edges, whereas
$N^\perp(\omega)$ is almost constant across the spectrum. Beyond
the spectral region the parameters $\xi^{\alpha\beta}(\omega)$ are
imaginary, and hence, the coherence length is finite, even in the
uniform chain. Remarkably, upon variation of the parameter $\Phi$
the two lengths $N^\parallel(\omega)$ and $N^\perp(\omega)$ evolve
differently: the first decreases while the second grows with
disorder. Regarding the exact diagonalization approach, the
coherence length can be extracted directly from envelope of the
disorder-averaged real-space spectral density in accordance with
Eq.~(\ref{NCohDef}). Unlike the case of DOS and absorption, much
larger statistical error is present owing to the fact that the
quantity to be averaged (i.e., a tensorial component of the
spectral density) depends on both energy and the real-space
variables. Here one should keep small the width of the energy
integration interval [$\delta \omega \approx 1/400$ of the total
bandwidth in our simulations] to ensure that the additional
decoherence due to interference of waves separated by energies
smaller than $\delta\omega$ does not exceed the disorder-induced
effect. If $v(\omega)$ is the excitation group velocity the last
condition can be expressed as $v(\omega) / \delta \omega \gg
N^\parallel(\omega)$. This inequality is not easy to fulfill
around the band center [$N^\parallel(\omega)$ is large] and around
the band edges [$v(\omega)$ is small]. Apart from mentioned the
numerical procedure suffers from the finite-size quantization
taking place as the coherence length approaches the total length
of the chain.

\begin{figure}
\centering
\includegraphics[width=5.25cm]{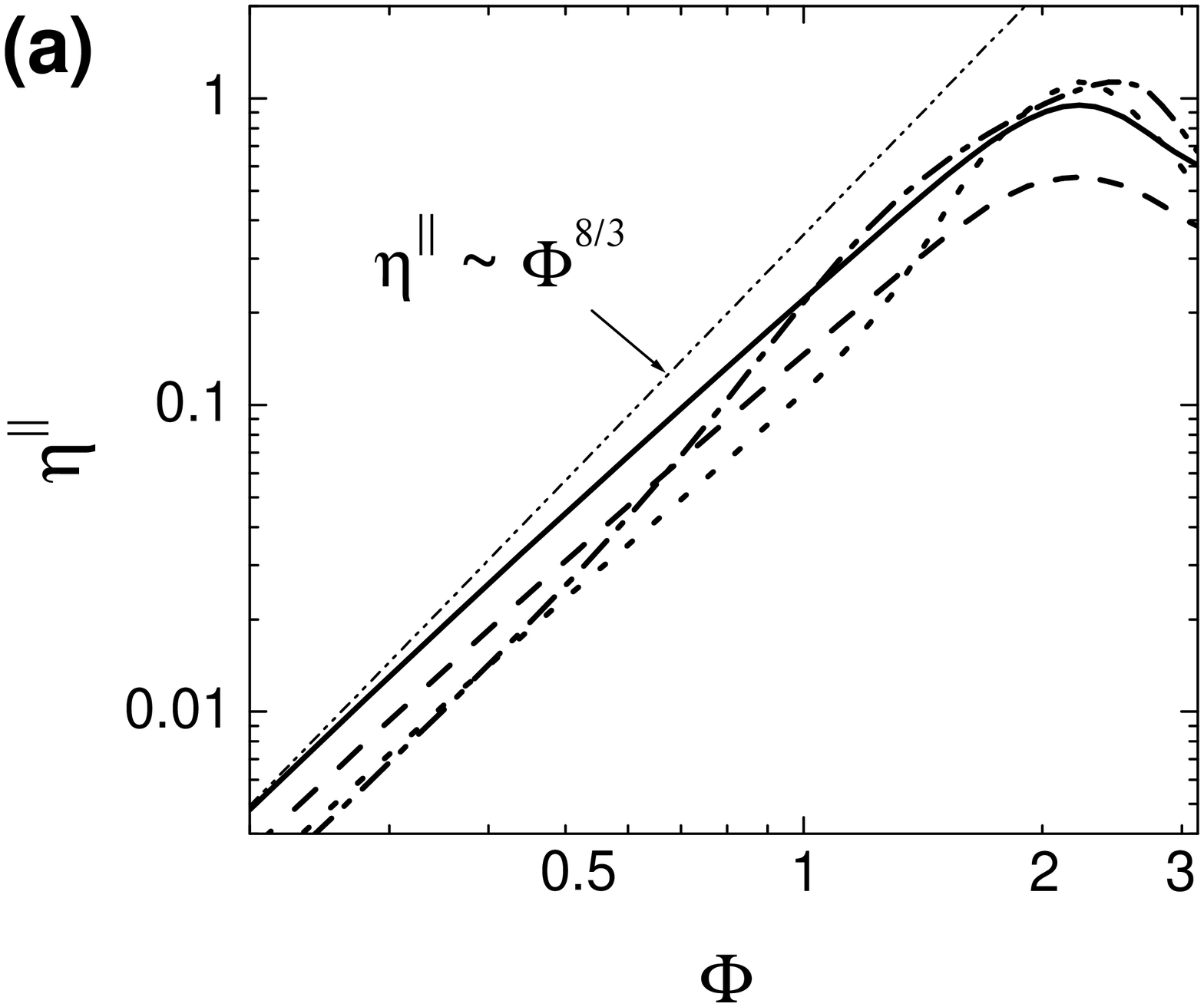}
\includegraphics[width=5.25cm]{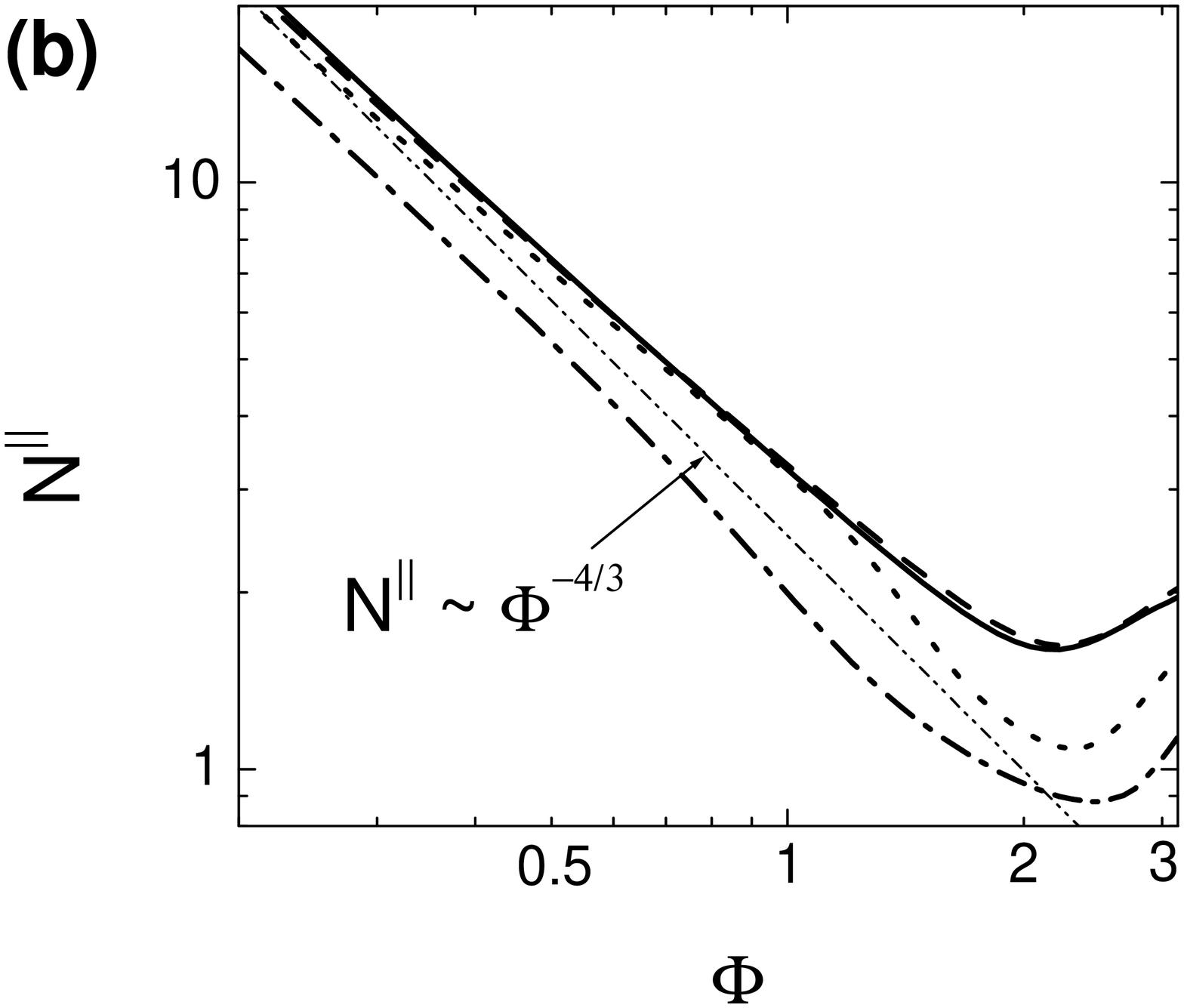}
\includegraphics[width=5.25cm]{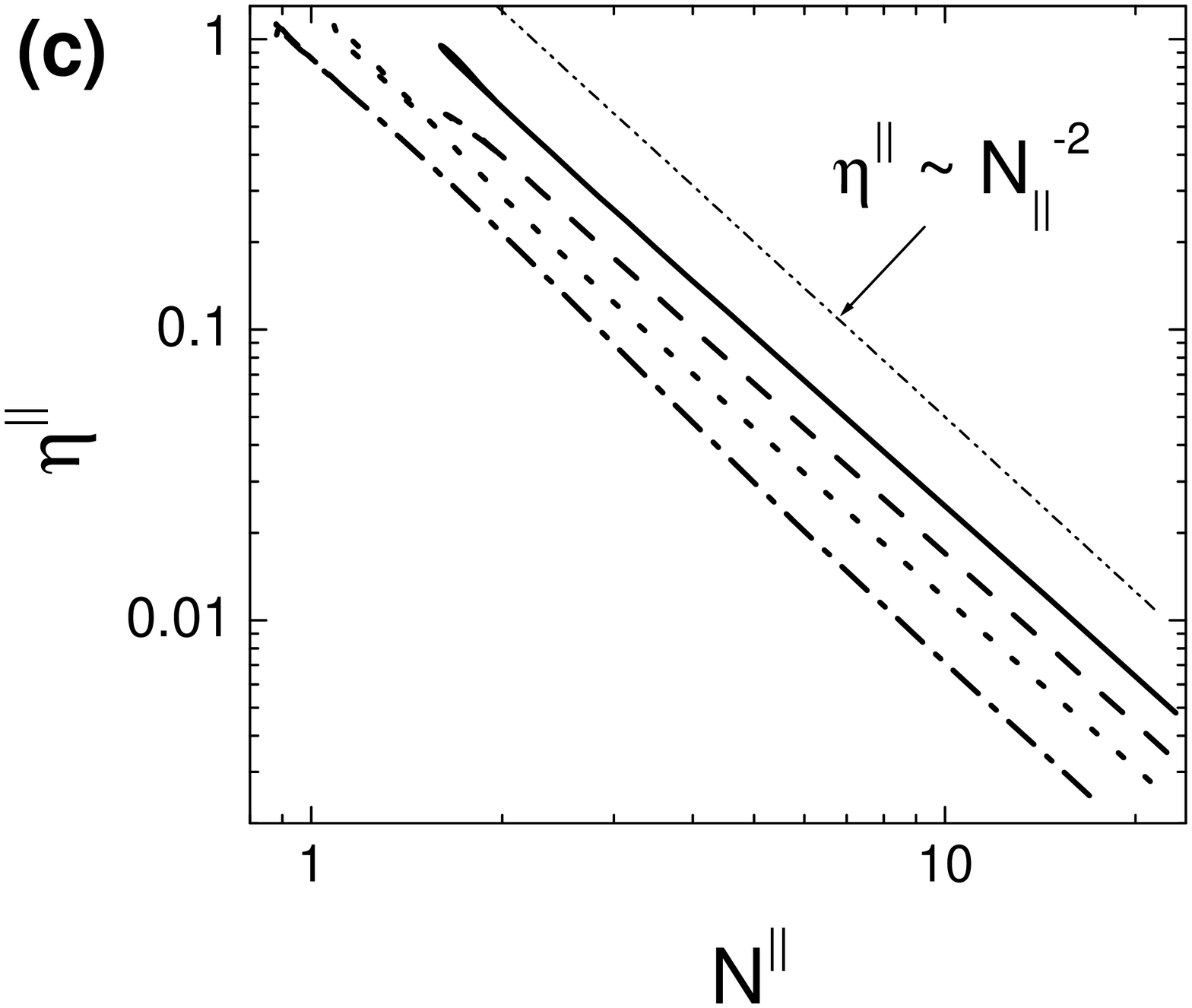}
\caption{\label{fig:Scaling} Dependencies of the absorption
linewidth (a) and the exciton coherence length at the resonance
(b) on the disorder parameter $\Phi$. (c) The same data is plotted
in the coordinates ``absorption linewidth --- coherence length".
The parameter $\theta$ is 0.26 (solid line), 0.52 (dashed line),
0.65 (dotted line), 0.78 (dash-dotted line).}
\end{figure}

For the model under consideration we have also verified the
accuracy of the low-disorder analytic solutions found in
Sec.~\ref{sec:WeakDis}. Since no diagonal disorder is present the
centered moments $a_2$ and $b_2^\alpha$ of the distribution
vanish, and the only parameters which control the strength of
disorder are $c_2^\parallel \approx \Phi^4/45$ and $c_2^\perp
\approx \Phi^2/3$. To evaluate these constants we have used
Eq.~(\ref{ABCDef}) and retained only the leading terms in the
limit of small $\Phi$. In the magnitude of the averaged dipole
moment one has to keep also the next-to-the-leading term, that
results to the estimate $|p| \approx 1 - \Phi^2/6$. In
Fig.~\ref{fig:Chi} (panels with $\Phi = 0.6$) the analytic
low-disorder absorption spectra are confronted with those
calculated via direct solution of the BEB-CPA equations and by
exact diagonalization. For polarization perpendicular to the
preferred orientation of dipoles the spectral weight of the
band-edge singularities (located at $\pm\zeta = \pm 2(1 -
3\cos^2\theta)$ for NN coupling) is negligible compared to the
integrated spectral weight. To approximate such wide-bandwidth
absorbance one can use the iterative solution derived in
Sec.~\ref{sec:Iter} which results to expression $\Im\bar
\chi^\perp(\omega) \approx {\cal N} c^\perp_2 /
\sqrt{\zeta^2-\omega^2}$. In contrast, the component
$\Im\bar\chi^\parallel(\omega)$, with most of the oscillator
strength concentrated around the lower band edge, should be
calculated using the approach of Sec.~\ref{sec:Scaling}. In
particular, taking into account that the disorder parameter
(\ref{DisorderCombined}) is given by $\Delta \sim \Phi^2$, the
scalings (\ref{EtaScal}) and (\ref{NScal}) read $\eta^\parallel
\sim \Phi^{8/3}$, $N^\parallel \sim \Phi^{-4/3}$, respectively.
The linewidth $\eta^\parallel$ and the exciton coherence length
$N^\parallel$ at the absorption maximum extracted from the
numerical solution of the BEB-CPA equations are plotted against
$\Phi$ in Fig.~\ref{fig:Scaling}(a,b). The presented profiles
confirm the fulfillment of the derived scaling laws. Moreover, as
illustrates Fig.~\ref{fig:Scaling}(c), the universal scaling
(\ref{NETAScal}) is observed in the whole range of
$\eta^\parallel$ and $N^\parallel$ accessible by varying $\Phi$
from $0$ to $\pi$, even though the two above dependencies hold
only for $\Phi$ sufficiently close to zero.

\subsection{Experimental relevance}

In a typical experiment one deals not with a single molecular
chain but rather with a macroscopic ensemble of such objects. The
treatment reported here is applicable only if the coupling between
molecules of distinct aggregates is weaker than the
disorder-induced broadening in a single chain. The broadening
mechanisms not related to the considered types of disorder can be
accounted for via a phenomenological on-site dephasing rate
$\eta$, e.g., due to exciton-phonon scattering. The results of the
BEB-CPA theory can be straightforwardly extended to include this
case by adding imaginary contribution $i \eta$ to the complex
energy variable $z$ in all relevant functions.

In most of the samples accessible in experiment, such as frozen
solutions containing molecular aggregates, because of the
randomness in the chain's orientations one is not able to measure
the individual spatial components of the microscopic
(single-chain) polarizability tensor. Nevertheless, the available
types of surface-deposited films\cite{LangBlodg} with strictly
in-plane alignment or recently synthesized bulk materials with
smectic-like\cite{Smectic} alignment of aggregates still allow for
independent observation of certain spatial components of the
single-chain polarizability. With a polarization-resolved
spectroscopic study of such systems\cite{PolarRes} one can probe
the orientational disorder within a single aggregate. In
particular, a material obtained by stacking the chains described
in Fig.~\ref{Fig2} preserving perfectly parallel alignment of the
axes and planes represents a biaxial medium. The optical axes are
given by the directions $e^\alpha_\parallel$, $e^\alpha_\perp$,
with the corresponding polarizabilities proportional to
$\bar\chi^\parallel(\omega)$ and $\bar\chi^\perp(\omega)$, and the
one orthogonal to them, along which the optical response vanishes.

\section{Conclusions}

In summary, using BEB CPA we have investigated the linear optical
properties of a disordered molecular aggregate with random
single-molecule excitation energies and transition dipoles. The
advantages of the used self-consistent approximation are its
analyticity, exactness in the low-disorder limit and the
capability to provide not only the DOS and absorption spectra but
also the coherence length. To the best of our knowledge, the last
quantity has never been addressed in excitonic systems with the
coherent-potential approximation. Whereas the DOS is on-site
characteristics while the absorption spectrum is related to the
spectral density at momentum $k=0$, the calculation of coherence
length requires knowledge of the spectral density at a generic
$k$. That is, the last quantity carries essentially new
information on the single-particle properties of the system with
respect to that contained in the first two.

The interesting feature of the considered problem is the tensorial
structure of equations brought about by orientational disorder in
the transition dipoles. Such type of disorder leads to
redistribution of absorbance between different polarization
components with quite distinct behavior of the corresponding
coherence lengths.

\begin{acknowledgments}

We thank Professor F.~Bassani for helpful discussions and
encouragement and Dr D.M. Basko for suggestions on improving the
manuscript. Financial support from U.V.O.-R.O.S.T.E. UNESCO, the
European Union RTN-HYTEC and MIUR (PRIN-2001 ``Advanced hybrid
heterostructures'') is gratefully acknowledged.

\end{acknowledgments}

\appendix

\section{BEB-CPA equations for orientational disorder with NN and NNN coupling \label{BEBExpl}}

In the case of box probability distribution the configurational
averaging in the self-consistency equation (\ref{CPA1Matr}) can be
performed analytically leading to
\begin{subequations}
\begin{eqnarray}
\label{ModelSC11} \ds (\Sigma^\parallel - \Sigma^\perp) \bar
\Gamma^\parallel_{nn} = q \sqrt{\frac{z -
\Sigma^\perp}{z-\Sigma^\parallel}}  - 1,
\\
\label{ModelSC22} \ds (\Sigma^\perp - \Sigma^\parallel) \bar
\Gamma^\perp_{nn} = q \sqrt{\frac{z -
\Sigma^\parallel}{z-\Sigma^\perp}} - 1,
\end{eqnarray}
\end{subequations}
where function $q(z)$ is to be found from
\begin{equation}
\label{ModelQ} \tan( q \Phi ) =
\sqrt{\frac{z-\Sigma^\perp}{z-\Sigma^\parallel}} \tan \Phi.
\end{equation}
From Eq.~(\ref{SigmaSimple}) the projected components of the
self-energy can be expressed as
\begin{equation}
\label{ModelS} \Sigma^\parallel = \frac1{\gamma^\parallel} -
\frac1{\bar \Gamma^\parallel_{nn}}, \qquad \Sigma^\perp =
\frac1{\gamma^\perp} - \frac1{\bar \Gamma^\perp_{nn}}.
\end{equation}

Performing momentum integration in Eq.~(\ref{CPA2Matr}) one
relates the nontrivial components of the disorder-averaged on-site
GT to the coherent-potential GT as
\begin{equation}
\label{ModelGnn} \bar\Gamma^\parallel_{nn} = \gamma^\parallel
f\bigl( (1-3\cos^2\theta) \gamma^\parallel \bigr), \quad
\bar\Gamma^\perp_{nn} = \gamma^\perp f( \gamma^\perp ).
\end{equation}
The function $f(x)$ entering these formulas is given by
$$
\frac1{\sqrt{1 - 4 x^2}},
$$
and
$$
\begin{array}{r}
\ds \frac2{\sqrt{4+9x}}\left\{ \left[ \left(\sqrt{8x} +
\sqrt{4+9x} \right)^2 - 2x \right]^{-1/2} \right.
\\
\ds \left. + \left[ \left(\sqrt{8x} - \sqrt{4+9x} \right)^2 - 2x
\right]^{-1/2} \right\},
\end{array}
$$
for the cases of NN and NNN coupling, respectively.

For every disorder parameter $\Phi$ and tilt $\theta$ the equation
presented in this Appendix can be solved numerically for
$\gamma^\parallel(z)$ and $\gamma^\perp(z)$. The example of such
solution is shown in Fig.~\ref{fig:Self}.

\section{\label{app:Sing} Flat-band singular spectra}

In this appendix we study the single-particle properties of the
system described in Sec.~\ref{sec:SimpleModel} when the angle
$\theta$ formed by the plane of the transition dipoles with the
chain's direction is such that $\cos^2\theta = 1/3$ (i.e., $\theta
= 0.955...$). The disorder-free situation for this value of
$\theta$ can be considered as an intermediate case between that of
$J$ and $H$ aggregate, because the transfer energy
(\ref{TransferTerm}) generated by the dipolar coupling
(\ref{DDKernel}) vanishes. Neglecting higher multipole
contributions to the transfer energy as well as all other
broadening mechanisms which go beyond the present model, let us
concentrate on the effect of orientational disorder to produce a
finite excitonic bandwidth. [Considering instead the diagonal
disorder alone results to a trivial physics: the off-diagonal part
of the Hamiltonian is always zero and the molecules remain
uncoupled.] Our aim is to find using BEB CPA the asymptotic form
of the GT around $z=0$, where the DOS and certain components of
polarizability tensor have a singularity. [Such singularity due to
specific arrangement of coupling and disorder should not to be
confused with the Dyson singularity\cite{Dyson}, not accessible
within BEB CPA]. For simplicity we shall consider only the case of
NN coupling.

\begin{figure}
\centering
\includegraphics[width=8cm]{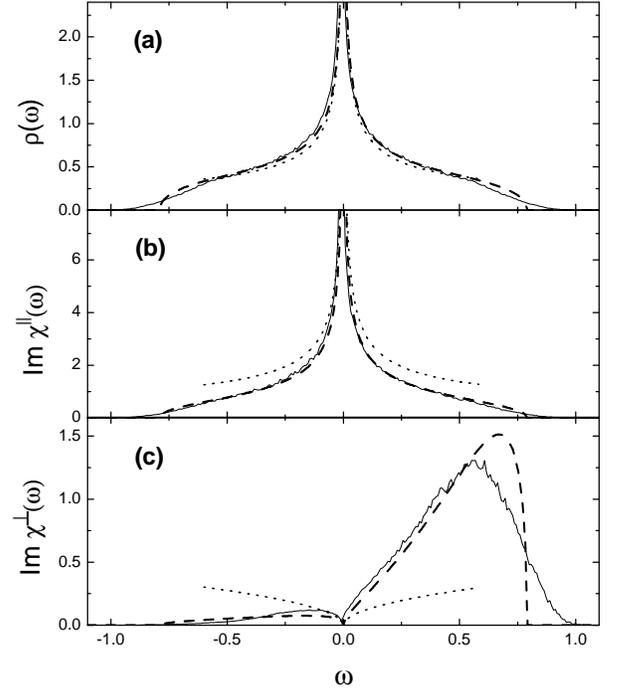}
\caption{\label{fig:Singul} (a) Exciton DOS and (b), (c)
absorption spectra in the case $\cos^2\theta = 1/3$. The BEB-CPA
results (dashed lines) are confronted with ones of the exact
diagonalization (solid lines) and the derived asymptotic
expressions (dotted lines). The disorder parameter $\Phi$ is equal
to $0.9$ for all curves.}
\end{figure}

From Eq.~(\ref{ModelGnn}) with $\cos^2\theta = 1/3$ it follows
that $\bar\Gamma^\parallel_{nn}(z) = \gamma^\parallel(z)$.
According to Eqs.~(\ref{ModelS}) the corresponding component of
the BEB-CPA self-energy is zero, $\Sigma^\parallel(z) = 0$. On the
other hand $\Sigma^\perp(z)$ is nonzero at any finite $z$. Taking
this into account it is easy to conclude that the auxiliary
function defined by formula (\ref{ModelQ}) behaves as $q(z) \to
\pi / (2 \Phi)$ upon $z \to 0$. Applying the same limit to
Eqs.~(\ref{ModelSC22}) we get
\begin{equation}
\label{SigmaAs} \bar\Gamma^\perp_{nn}(z) \sim
-\frac1{\Sigma^\perp(z)} \sim -\frac{i}2,
\end{equation}
\begin{equation}
\gamma^\perp(z) \sim -\frac{i \Phi}{\pi} \sqrt{\frac{2 i}z}.
\end{equation}
Combining formula (\ref{SigmaAs}) with Eq.~(\ref{ModelSC11}) we
obtain the asymptotics of the remaining unknown functions at $z
\to 0$:
\begin{equation}
\bar\Gamma^\parallel_{nn}(z) = \gamma^\parallel(z) \sim -\frac{i
\pi}{4\Phi} \sqrt{\frac{2i}z}.
\end{equation}
It turns out that, irrespectively on the strength $\Phi$ of
orientational disorder, the component
$\bar\Gamma^\parallel_{nn}(z)$ of the disorder-averaged on-site GT
is singular as $z \to 0$, i.e., at the energy of the bare
molecular transition. Strictly speaking, upon introducing some
degree of diagonal disorder or a finite inelastic dephasing rate
$\eta$ such singularity would be suppressed. Nevertheless, one can
still apply the derived asymptotic formulas if the absolute value
of $z$ is much larger than $\eta$ but smaller than the total
bandwidth generated by orientational disorder.

The exciton DOS, related to the on-site GT found above by formula
(\ref{OrientDOS}), shows an analogous singularity around $\omega =
0$:
\begin{equation}
\bar \rho(\omega) \sim \frac1{4\Phi} \frac1{\sqrt{|\omega|}}.
\end{equation}
As concerns the absorption spectra, for the two essential
polarization directions these are characterized by completely
distinct behavior:
\begin{equation}
\Im \bar \chi^\parallel(\omega) \sim \frac{{\cal N} \pi}{4\Phi}
\frac1{\sqrt{|\omega|}}, \quad \Im \bar \chi^\perp(\omega) \sim
\frac{{\cal N} \pi}{8\Phi} \sqrt{|\omega|}.
\end{equation}
The DOS and the absorption profiles calculated via numerical
solution of the BEB-CPA equations and ones of exact
diagonalization are confronted with the derived asymptotic
formulas in Fig.~\ref{fig:Singul}. All approaches agree well close
to the band center. Noticeably, unlike the usual case
$\cos^2\theta \ne 1/3$, the absorption component $\Im
\bar\chi(\omega)$ is symmetric with respect to the central energy
$\omega = 0$ (still only the NN coupling is assumed).

Regarding the exciton coherence length, since
$\vartheta^\parallel_k = 0$ it follows that $N^\parallel(z) = 0$
for every $z$. In other words, for polarization along the
preferred orientation of the dipoles the disorder-averaged
spectral density depends on the intermolecular separation as $\Im
\bar\Gamma^\parallel_{nm}(z) \sim \delta_{nm}$, i.e., we encounter
completely incoherent optical response of the molecules. For the
orthogonal polarization in the vicinity of $z = 0$ we have
\begin{equation}
\xi^\perp(z) \sim \frac{\pi}2 + \frac{i\pi}{2\Phi}
\sqrt{\frac{z}{2i}},
\end{equation}
\begin{equation}
N^\perp(\omega) \sim \frac{4\Phi}{\pi} \frac1{\sqrt{|\omega|}}.
\end{equation}
As a result the coherence length $N^\perp(\omega)$ is finite
within the band but diverges at its center. In the vicinity of
$\omega = 0$ the corresponding real-space spectral density behaves
as $\Im\bar\Gamma^\perp_{nm}(0) \sim \cos\bigl( \pi |n-m|/2
\bigr)$, i.e., coincides with the center-of-the-band spectral
density of a finite-bandwidth uniform chain.


\begin{thebibliography}{}
\bibitem{Kozlov} As an example of the opposite behavior we can mention the nonexponential
localization of the band-center wave function in a
nearest-neighbor chain with off-diagonal disorder [see
Ref.~\onlinecite{KozlovRef}].
\end{thebibliography}
\end{document}